\documentclass[prb,twocolumn,superscriptaddress,floatfix,letterpaper]{revtex4}
\pdfoutput=1

\usepackage{graphicx}
\usepackage{amsmath}
\usepackage{amstext}
\usepackage{amssymb}
\usepackage{mathrsfs}
\usepackage{amsfonts}
\usepackage{amsbsy} 
\usepackage{dcolumn}
\usepackage{bm}
\usepackage{multirow}
\usepackage{amsthm} 

\usepackage{color}
\usepackage[colorlinks,citecolor=blue]{hyperref}
\definecolor{red}{rgb}{1,0,0}
\definecolor{blue}{rgb}{0,0,1}
\definecolor{dblue}{rgb}{0,0,0.4}
\definecolor{green}{rgb}{0,1,0}
\definecolor{black}{rgb}{0,0,0}
\definecolor{white}{rgb}{1,1,1}

\definecolor{brn}{rgb}{.8,.4,.0}
\definecolor{redo}{rgb}{1,.5,.0}
\definecolor{ddgrn}{rgb}{0,0.4,0}
\definecolor{dgrn}{rgb}{0,0.55,0}
\definecolor{dbl}{rgb}{0,0,0.5}

\usepackage[bbgreekl]{mathbbol}
\usepackage{amscd}

\newcommand{\Z}{\mathbb{Z}}

\newcommand{\R}{\mathbb{R}}

\newcommand{\N}{\mathbb{N}}

\renewcommand{\v}[1]{\boldsymbol{#1}} 
\renewcommand{\t}[1]{\tilde{#1}} 
\newcommand{\ii}{\hspace{1pt}\mathrm{i}\hspace{1pt}}
\newcommand{\ee}{\hspace{1pt}\mathrm{e}}

\newcommand{\<}{\langle} 
\renewcommand{\>}{\rangle} 

\newcommand{\Ref}[1]{Ref.~\onlinecite{#1}}
 
\newcommand{\eq}[1]{(\ref{#1})} 
\newcommand{\eqn}[1]{eqn.~(\ref{#1})}

\newcommand{\Tr}{{\rm Tr}}

\newcommand{\prt}{\partial}

\newcommand{\ie}{{\it i.e.~}} 
 
\newcommand{\etc}{{\it etc~}}

\newcommand{\cH}{ {\cal H} }

\newcommand{\cV}{ {\cal V} }

\newcommand{\al}{\alpha} 
\newcommand{\bt}{\beta} 
\newcommand{\del}{\delta} 
 
\newcommand{\eps}{\epsilon}

\newcommand{\la}{\lambda} 
\newcommand{\La}{\Lambda}

\newcommand{\si}{\sigma}

\newcommand{\bpm}{\begin{pmatrix}}
\newcommand{\epm}{\end{pmatrix}}
\newcommand{\bmm}{\begin{matrix}}
\newcommand{\emm}{\end{matrix}}
 
\let\baraccent=\= 
\renewcommand{\=}[1]{\stackrel{#1}{=}} 

\newcommand{\ad}{\text{ad}}
\newcommand{\frc}[2]{{{#1}/{#2}}}

\begin{document}

\title{ 
Simple-current algebra constructions of 2+1D topological orders
}

\author{Kareljan Schoutens}
\affiliation{Institute for Theoretical Physics, University of Amsterdam, Science Park 904, 1098 XH Amsterdam}

\author{Xiao-Gang Wen} 
\affiliation{Department of Physics, Massachusetts
Institute of Technology, Cambridge, Massachusetts 02139, USA}
\affiliation{Perimeter Institute for Theoretical Physics, Waterloo, Ontario,
N2L 2Y5 Canada} 

\date{August 5, 2015}

\begin{abstract} 
Self-consistent (non-)abelian statistics in 2+1D are classified by modular tensor categories (MTC).  In recent works, a simplified axiomatic approach to MTCs, based on fusion coefficients $N^{ij}_k$ and spins $s_i$, was proposed. A numerical search based on these axioms led to a list of possible (non-)abelian statistics, with rank up to $N=7$. However, there is no guarantee that all solutions to the simplified axioms are consistent and can be realised by bosonic physical systems.  In this paper, we use simple-current algebra to address this issue. We explicitly construct many-body wave functions, aiming to realize the entries in the list (\ie realize their fusion coefficients $N^{ij}_k$ and spins $s_i$).  We find that all entries can be constructed by simple-current algebra plus conjugation under time reversal symmetry.  This supports the conjecture that simple-current algebra is a general approach that allows us to construct all (non-)abelian statistics in 2+1D. It also suggests that the simplified theory based on $(N^{ij}_k,s_i)$ is a classifying theory at least for simple bosonic 2+1D topological orders (up to invertible topological orders).

\end{abstract}

\maketitle

{\small \setcounter{tocdepth}{2} \tableofcontents }

\section{Introduction}

We know that symmetry breaking orders\cite{L3745,LanL58} are described by group
theory, which allows us to classify all different symmetry breaking orders.  It is then natural to ask what mathematical theory classifies topological orders\cite{Wrig,KW9327}, which are beyond symmetry breaking orders. One proposal is to use the properties of topological excitations (such as their (non-)abelian statistics) to classify topological orders. This has led to the proposal that $d+1D$ bosonic topological orders can be classified by unitary $(d+1)$-categories with one object.\cite{KW1458,KWZ1590} In particular, unitary $(2+1)$-categories with one object are modular tensor categories (MTC), leading to the proposal that 2+1D bosonic topological orders are classified by MTCs.\cite{W8951,GKh9410089,BK01,K062,Bon07,
RSW0777,Wang10,onlineMTC15,W150605768} Such a classification is up to invertible topological orders, which have no nontrivial topological excitations.\cite{KW1458,F1478} 

\subsection{Simplified axiomatic approach}

The papers \Ref{RSW0777,W150605768} have formulated a simplified axiomatic approach to MTCs. This approach is based on fusion coefficients $N^{ij}_k$ and spins $s_i$; it does not explicitly involve more involved data such as $R$- and $F$-matrices. The simplified axioms were used for a numerical search of simple MTCs, which led to a list of possible bosonic topological orders in 2+1D, with rank up to $N=7$ (see Tables \ref{toplst} - \ref{toplst7}). 

For certain special types of topological orders, the classification can be
described  by simpler theories.  For example,  topological orders with gappable
edge for 2+1D interacting bosonic systems can be classified by unitary fusion
categories (UFC).\cite{LWstrnet,CGW1038} For 2+1D bosonic/fermionic topological
orders (with gappable or un-gappable edge) that have only abelian statistics,
we can use integer $K$-matrices to classify them\cite{WZ9290} and use the
following $U(1)$ Chern-Simons theory to describe
them\cite{BW9045,R9002,FK9169,WZ9290,BM0535,KS1193,Wtoprev}
\begin{align}
\label{csK}
 {\cal L}= \frac{K_{IJ}}{4\pi} a_{I\mu} \prt_\nu a_{J\la}\eps^{\mu\nu\la} .
\end{align}
Such an effective theory can be realized by a multi-layer fractional quantum
Hall state:
\begin{align}
\label{wavK}
\prod_{I;i<j} (z_i^I-z_j^I)^{K_{II}}
\prod_{I<J;i,j} (z_i^I-z_j^J)^{K_{IJ}}
\ee^{-\frac14 \sum_{i,I} |z_i^I|^2}.
\end{align}
When the diagonal elements $K_{II}$ are all even, the $K$-matrices classify
2+1D bosonic abelian topological orders.  When some diagonal elements
$K_{II}$ are odd, the $K$-matrices classify 2+1D fermionic abelian
topological orders.

The list produced in \Ref{RSW0777,W150605768} gives solutions to the simplified axioms for MTCs - as such it describes possible self-consistent (non-)abelian statistics in 2+1D. However, there is no guarantee that all solutions are indeed consistent and can be realized by many-boson wavefunctions.  

\subsection{Simple-current algebra constructions}

In this paper, we pursue a constructive (rather than axiomatic) approach to bosonic topological orders in 2+1D. We use simple-current algebra to construct and classify such orders, and demonstrate that simple-current algebras can produce all orders listed in Tables \ref{toplst} - \ref{toplst7}.  

It is well-known that correlation functions in conformal field theory (CFT) can be used to construct many-body wave functions\cite{MR9162,BW9215,WW9455,LWW1024,WW0808,WW0809,BW0932} that realize
topological orders in 2+1D. In this paper we use these ideas to arrive at many-boson wavefunctions for bosonic topological orders. The main building blocks for our constructions are a set of CFT simple currents
\begin{align}
 \psi_I, \ \ \ \ I=1,\cdots,M \ .
\end{align}
We combine these with scalar field vertex operators to define 
\begin{align}
 c_I=\psi_I \ee^{\ii \sum_\mu k^I_\mu \phi^\mu}= \psi_I \ee^{\ii \v k^I\cdot \v \phi}
\end{align}
and construct bosonic wave functions as
\begin{align}
P(\{z_i^I\})
=\lim_{z_\infty\rightarrow\infty}\langle
V(z_\infty)\prod_{i,I}
c_I(z_i^I)
\rangle .
\end{align}
We refer to section \ref{FQHVA} for details and further explanation.

In this paper and in \Ref{LWW1024}, we like to stress that it is misleading to state that CFT as such classifies topological orders. It is really simple-current algebra that can be used to classify 2+1D  topological orders.  
In this paper, we show how to calculate the fusion coefficients $N^{ij}_k$ and spins $s_i$ of the topological excitations from simple-current algebra. This allows us to recover all entries in the Tables
\ref{toplst} - \ref{toplst7} using simple-current algebra. 

The consistency of the MTC axioms of \Ref{RSW0777,W150605768} guarantees that all consistent orders are covered by lists such as those of Tables \ref{toplst} - \ref{toplst7}. In that sense those lists are an upper bound to the actual list of all consistent orders. The orders coming out of simple-current algebra constructions are consistent by construction - they thus establish a lower bound to the list of all consistent orders. In all cases considered in this paper, the two bounds agree, allowing us to conclude that both the simplified axiomatic approach and the simple-current algebra constructive approach give complete results.

\def\arraystretch{1.25} \setlength\tabcolsep{3pt} 
\begin{table*}[tb] 
\caption{ A list of 35 bosonic topological orders in 2+1D with rank $N=1,2,3,4$
and with max$(N^{ij}_k)\leq 3$.  All $N\leq 4$ orders have max$(N^{ij}_k)=1$.
The entries in blue are composite topological orders that can be obtained by
stacking lower rank topological orders. The first column is the rank $N$ and
the central charge $c$ (mod 8).  The second column is the topological
entanglement entropy $S_\text{top}=\log_2 D$, $D=\sqrt{\sum_i d_i^2}$.  The
quantum dimensions of the topological excitations in the third column are
expressed in terms of $\zeta_n^m=\frac{\sin[\pi(m+1)/(n+2)]}{\sin[\pi/(n+2)]}$.
The fourth column are the spins of the corresponding topological excitations.
By `type $(X_l,k)$' we indicate a correspondence to affine Kac-Moody current
algebra $X_l^{(1)}$ at level $k$, and $(X_l,k)_{1 \over q}$ indicate simple-current
reductions of Kac-Moody current algebra.
} 
\vskip 2mm
\label{toplst} 
\centering
\begin{tabular}{ |c|c|c|c|c||c|c|c|c|c| } 
\hline 
$N^B_c$ & $S_\text{top}$ & $d_1,d_2,\cdots$ & $s_1,s_2,\cdots$ & \text{type} &
$N^B_c$ & $S_\text{top}$ & $d_1,d_2,\cdots$ & $s_1,s_2,\cdots$ & \text{type} 
\\
 \hline 
 $1^B_{ 0}$ & $0$ & $1$ & $0$ & & & & & & 
 \\
 \hline 
 $2^B_{1}$ & $0.5$ & $1,1$ & $0, \frac{1}{4}$ & $U(1)_1$, $(A_1,1$) & 
 $2^B_{-1}$ & $0.5$ & $1,1$ & $0,-\frac{1}{4}$ & ($E_7,1$)
 \\
 $2^B_{\frc{14}{5}}$ & $0.9276$ & $1,\zeta_3^1$ & $0, \frac{2}{5}$ & ($G_2,1$), ($A_1,3)_{1 \over 2}$ & 
 $2^B_{-\frc{14}{5}}$ & $0.9276$ & $1,\zeta_3^1$ & $0,-\frac{2}{5}$ & ($F_4,1$), ($A_2,2)_{1 \over 3}$
 \\
 \hline 
$3^B_{ 2}$ & $0.7924$ & $1,1,1$ & $0, \frac{1}{3}, \frac{1}{3}$ & ($A_2,1$), $(A_1,4)_{1 \over 4}$ &
$3^B_{-2}$ & $0.7924$ & $1,1,1$ & $0,-\frac{1}{3},-\frac{1}{3}$ & ($E_6,1$) 
 \\
$3^B_{\frc{1}{2}}$ & $1$ & $1,1,\zeta_2^1$ & $0,\frac{1}{2}, \frac{1}{16}$ & $(B_8,1)$ &
$3^B_{-\frc{1}{2}}$ & $1$ & $1,1,\zeta_2^1$ & $0,\frac{1}{2},-\frac{1}{16}$ & ($B_7,1$), ($E_8,2$)
\\
$3^B_{\frc{3}{2}}$ & $1$ & $1,1,\zeta_2^1$ & $0,\frac{1}{2}, \frac{3}{16}$ & ($A_1,2$) &
$3^B_{-\frc{3}{2}}$ & $1$ & $1,1,\zeta_2^1$ & $0,\frac{1}{2},-\frac{3}{16}$ & ($B_6,1$)
\\
$3^B_{\frc{5}{2}}$ & $1$ & $1,1,\zeta_2^1$ & $0,\frac{1}{2}, \frac{5}{16}$ & ($B_2,1$) &
$3^B_{-\frc{5}{2}}$ & $1$ & $1,1,\zeta_2^1$ & $0,\frac{1}{2},-\frac{5}{16}$ & ($B_5,1$)
\\
$3^B_{\frc{7}{2}}$ & $1$ & $1,1,\zeta_2^1$ & $0,\frac{1}{2}, \frac{7}{16}$ & ($B_3,1$) &
$3^B_{-\frc{7}{2}}$ & $1$ & $1,1,\zeta_2^1$ & $0,\frac{1}{2},-\frac{7}{16}$ & ($B_4,1$)
\\
$3^B_{\frc{8}{7}}$ & $1.6082$ & $1,\zeta_5^1,\zeta_5^2$ & $0,-\frac{1}{7}, \frac{2}{7}$ & ($A_1,5)_{1 \over 2}$ &
$3^B_{-\frc{8}{7}}$ & $1.6082$ & $1,\zeta_5^1,\zeta_5^2$ & $0, \frac{1}{7},-\frac{2}{7}$ & $(A_4,2)_{1 \over 5}$ 
\\
\hline 
$4^{B,a}_{ 0}$ & $1$ & $1,1,1,1$ & $0, 0,0,\frac{1}{2}$ & ($D_8,1$) &
\color{blue} $4^{B,b}_{ 0}$ &\color{blue}  $1$ &\color{blue} $1,1,1,1$ &\color{blue}  $0, 0, \frac{1}{4},-\frac{1}{4}$ &
\\
$4^B_{ 1}$ & $1$ & $1,1,1,1$ & $0, \frac{1}{8}, \frac{1}{8},\frac{1}{2}$ & $U(1)_2$
 &
$4^B_{-1}$ & $1$ & $1,1,1,1$ & $0,-\frac{1}{8},-\frac{1}{8},\frac{1}{2}$ & ($D_7,1$)
 \\
\color{blue} $4^B_{ 2}$ &\color{blue}  $1$ &\color{blue}  $1,1,1,1$ &\color{blue}  $0, \frac{1}{4}, \frac{1}{4},\frac{1}{2}$ & &
\color{blue} $4^B_{-2}$ &\color{blue}  $1$ &\color{blue}  $1,1,1,1$ &\color{blue} $0,-\frac{1}{4},-\frac{1}{4},\frac{1}{2}$ & \color{blue} ($D_6,1$) 
\\
$4^B_{ 3}$ & $1$ & $1,1,1,1$ & $0, \frac{3}{8}, \frac{3}{8},\frac{1}{2}$ & ($A_3,1$) &
$4^B_{-3}$ & $1$ & $1,1,1,1$ & $0,-\frac{3}{8},-\frac{3}{8},\frac{1}{2}$ & ($D_5,1$)
\\
 $4^B_{4}$ & $1$ & $1,1,1,1$ & $0,\frac{1}{2},\frac{1}{2},\frac{1}{2}$ & ($D_4,1$), $(A_2,3)_{1 \over 9}$ &
 \color{blue} $4^{B,c}_{ 0}$ &\color{blue} $1.8552$ &\color{blue} $1,\zeta_3^1,\zeta_3^1,\zeta_3^1\zeta_3^1$ &\color{blue} $0, \frac{2}{5},-\frac{2}{5}, 0$ &
\\
 \color{blue} $4^B_{\frc{9}{5}}$ &\color{blue} $1.4276$ &\color{blue} $1,1,\zeta_3^1,\zeta_3^1$ &\color{blue} $0,-\frac{1}{4}, \frac{3}{20}, \frac{2}{5}$ &  \color{blue} ($A_1,3$) &
\color{blue} $4^B_{-\frc{9}{5}}$ &\color{blue} $1.4276$ &\color{blue} $1,1,\zeta_3^1,\zeta_3^1$ &\color{blue} $0, \frac{1}{4},-\frac{3}{20},-\frac{2}{5}$ \color{blue} &
\\
\color{blue} $4^B_{\frc{19}{5}}$ &\color{blue} $1.4276$ &\color{blue} $1,1,\zeta_3^1,\zeta_3^1$ &\color{blue} $0, \frac{1}{4},-\frac{7}{20}, \frac{2}{5}$ & &
\color{blue} $4^B_{-\frc{19}{5}}$ &\color{blue} $1.4276$ &\color{blue} $1,1,\zeta_3^1,\zeta_3^1$ &\color{blue} $0,-\frac{1}{4}, \frac{7}{20},-\frac{2}{5}$ &\color{blue} ($C_3,1$) 
\\
 \color{blue} $4^B_{\frc{12}{5}}$ &\color{blue} $1.8552$ &\color{blue} $1,\zeta_3^1,\zeta_3^1,\zeta_3^1\zeta_3^1$ &\color{blue} $0,-\frac{2}{5},-\frac{2}{5}, \frac{1}{5}$ & \color{blue} $(A_1,8)_{1 \over 4}$ &
\color{blue} $4^B_{-\frc{12}{5}}$ &\color{blue} $1.8552$ &\color{blue} $1,\zeta_3^1,\zeta_3^1,\zeta_3^1\zeta_3^1$ &\color{blue} $0, \frac{2}{5}, \frac{2}{5},-\frac{1}{5}$ &
\\
$4^B_{\frc{10}{3}}$ & $2.1328$ & $1,\zeta_7^1,\zeta_7^2,\zeta_7^3$ & $0, \frac{1}{3}, \frac{2}{9},-\frac{1}{3}$ & ($A_1,7)_{1 \over 2}$
& $4^B_{-\frc{10}{3}}$ & $2.1328$ & $1,\zeta_7^1,\zeta_7^2,\zeta_7^3$ & $0,-\frac{1}{3},-\frac{2}{9}, \frac{1}{3}$ & ($G_2,2$), $(A_6,2)_{1 \over 7}$ \\
\hline \end{tabular} 
\end{table*}

\newcommand{\hsc}[1]{h^\text{sc}_{#1}}
\newcommand{\thsc}[1]{\tilde{h}^\text{sc}_{#1}}

\section{Constructing topologically ordered state of a given non-abelian
type via a simple-current algebra}

\label{FQHVA}

In this paper, we will use charged particles in multilayer system under
magnetic field as a general and systematic way to realize 2+1D bosonic and
fermionic topologically ordered states.  We will assume all the particles are
in the first Landau level. Thus the many-body wave function has a form
\begin{align}
 \Psi(\{z_i^I\})=
 P(\{z_i^I\})\ee^{-\frac14\sum_{i,I}|z_i^I|^2},
\end{align}
where $i$ labels different particles, $I=1,\cdots,M$ labels different
layers, and $P(\{z_i^I\})$ is a (anti-)symmetric polynomial (under $z_i^I
\leftrightarrow z_j^I$), depending on the Bose or Fermi statistics of the
particles in the $I^\text{th}$ layer.  
In this paper, we are going to use such kind of systems to systematically
realize non-abelian topological orders for bosons and fermions.

\subsection{Symmetric polynomial $P(\{z_i^I\})$ as a correlation function
in a simple-current algebra}
\label{corr}

Let us consider a CFT generated by simple currents $c_I(z)$, $I=1,\cdots,M$.
By definition, simple currents are operators with unit quantum dimension.  The
correlation function of simple currents always has one conformal block.  If the
simple currents $c_I(z)$ are also bosonic with integer conformal dimension or
fermionic with half-integer conformal dimension, then we can use the
correlation function of the simple currents $c_I(z)$ to construct the
(anti-)symmetric polynomial $P(\{z_i^I\})$ \cite{MR9162,BW9215,WW9455,WWH9476}
\begin{align}
\label{PhiV}
P(\{z_i^I\})\propto \lim_{z_\infty\rightarrow\infty}\langle
V(z_\infty)\prod_{i,I}c_I(z_i^I)\rangle
\end{align}
where $V(z_\infty)$ represents a background to guarantee that the correlation
function be non-zero.  In fact $c_I(z)$ is related to the annihilation
operator for the bosons in the $I^{\rm th}$ layer.

Such an approach allows us to use different simple-current CFTs to
construct/label different many-boson wave function, which may correspond to
different 2+1D topologically ordered states.  For example, the Laughlin
wave function $P(\{z_i\})=\prod_{i<j} (z_i-z_j)^m$ can be
constructed this way by choosing a Gaussian CFT and choosing
\begin{align}
 c(z)= \ee^{\ii \sqrt m \phi(z)}
\end{align}
as the simple-current operator.
Here, the operator $\ee^{\ii a \phi(z)}$ has conformal dimension
$\frac{a^2}{2}$ and the following operator product expansion (OPE)
\begin{align}
& \ee^{\ii a\phi(z)}\ee^{\ii b\phi(w)}=
\nonumber\\
& \quad (z-w)^{ab} \ee^{\ii (a+b)\phi(w)}+O\Big((z-w)^{ab+1}\Big) .
\end{align}
In fact
\begin{align}
\prod_{i<j} (z_i-z_j)^m\propto \lim_{z_\infty\rightarrow\infty}\langle
\ee^{-\ii N\sqrt m \phi(z_\infty)}\prod_{i=1}^N\ee^{\ii \sqrt m \phi(z_i)}
\rangle .
\end{align}

To construct the abelian topologically ordered states described by the
$K$-matrix wave function \eq{wavK}, we can start with a Gaussian model
described by $\phi^\mu$ fields that have the following OPE
\begin{align}
\ee^{\ii l_\mu \phi^\mu(z)}\ee^{\ii l'_\mu\phi^\mu(w)}=(z-w)^{l_\mu G^{\mu\nu} l'_\nu} \ee^{\ii (l_\mu+l_\mu')\phi^\mu(w)}+\cdots
\end{align}
We see that $\ee^{\ii  k_\mu \phi^\mu(z)}$ has a conformal dimension 
\begin{align}
\frac12 \v k\cdot \v k \equiv 
\frac12 \sum_{\mu\nu} k_\mu G^{\mu\nu} k_\nu , 
\end{align}
where the inner product $\cdot$ is defined via $G^{\mu\nu}$. The metric $G^{\mu\nu}$ plays a crucial role, as a given choice of $G^{\mu\nu}$ leads to a specific set of momenta $k_\mu$ giving vertex operators with integral conformal dimension (or half-integral for fermionic theories), thereby setting the operator content of the theory. If we choose $c_I=
\ee^{\ii \v k^I\cdot \v\phi}
\equiv \ee^{\ii  \sum_\mu k^I_\mu \phi^\mu}
$, where $\v k^I=(k^I_1,k^I_2,\cdots)$ and $\v
\phi=(\phi^1,\phi^2,\cdots)$, we find that
\begin{align}
P(\{z_i^I\})= &
\prod_{I;i<j} (z_i^I-z_j^I)^{K_{II}} \prod_{I<J;i,j} (z_i^I-z_j^J)^{K_{IJ}}
\nonumber \\
\propto & \lim_{z_\infty\rightarrow\infty}\langle V(z_\infty)\prod_{i,I} c_I(z_i^I) \rangle ,
\end{align}
if the $\v k^I$ satisfy
\begin{align}
 K_{IJ} = \v k^I\cdot \v k^J.
\end{align}
In order to obtain an (anti-)symmetric polynomial $P(\{z_i^I\})$, we see that
$K_{IJ}$ must be integer.

Now, we are ready to construct topologically ordered states of a given
non-abelian type.  Let us consider a simple-current CFT generated by a set of
simple currents
\begin{align}
 \psi_I, \ \ \ \ I=1,\cdots,M \ .
\end{align}
We assume that the $\psi_I$ have finite orders described by an integer matrix
$\v n=(n_{JI})$:
\begin{align}
\label{psin}
\prod_J (\psi_J)^{n_{JI}}=1,\ \ \forall \ I .
\end{align}
Now we choose 
\begin{align}
 c_I=\psi_I \ee^{\ii \sum_\mu k^I_\mu \phi^\mu}= \psi_I \ee^{\ii \v k^I\cdot \v \phi}
\end{align}
to construct the wave function as
\begin{align}
P(\{z_i^I\})
=\lim_{z_\infty\rightarrow\infty}\langle
V(z_\infty)\prod_{i,I}
c_I(z_i^I)
\rangle .
\end{align}
But in this case, in order to obtain an (anti-)symmetric polynomial $P(\{z_i^I\})$,
\begin{align}
K_{IJ}\equiv \v k^I\cdot \v k^J
= \sum_{\mu\nu}  k^I_\mu G^{\mu\nu}k^J_\nu 
\end{align}
may not be integer. In fact, introducing
\begin{align}
c_{\vec a} = \prod_I c_I^{a_I} 
= \ee^{\ii\sum_{I,\mu} a_I  k^I_\mu  \phi^\mu } \prod_I \psi_I^{a_I} 
\end{align}
and noticing that $c_{\vec a}$ and $c_{\vec b}$ must be mutually local for
any integer vectors $\vec a$ and $\vec b$, we find that $k^I_\mu$ must satisfy
\begin{align}
\label{Khhh}
& \sum_{IJ\mu\nu}  a^I k^I_\mu G^{\mu\nu}k^J_\nu  b^J -   \hsc{\vec a} -\hsc{\vec b} +\hsc{\vec a+\vec b} 
\nonumber\\
& = \sum_{IJ}  a^I K_{IJ}  b^J -   \hsc{\vec a} -\hsc{\vec b} +\hsc{\vec a+\vec b}
\in \N
\end{align}
for any positive integer vector $\vec a$ and $\vec b$ (\ie $a_I \in \N$ and
$b_I \in \N$).  Here $\N=\{0,1,2,\cdots\}$ and $\hsc{\vec a}$ is the conformal
dimension of $\psi_{\vec a}\equiv \prod_I \psi_I^{a_I}$.  Since the $\hsc{\vec
a}$ are rational numbers, in general, $K_{IJ}$ are also rational numbers.  We
see that, starting from a simple-current CFT, we can construct all the
2+1D topological orders of a given non-abelian type, by finding all the
$K$-matrices that satisfy the conditions \eq{Khhh}.

\begin{table*}[tb] 
\caption{A list of 10 bosonic rank $N=5$ topological orders in 2+1D with max$(N^{ij}_k)\leq 3$. 
The orders $5^B_{\pm \frc{18}{7}}$ have max$(N^{ij}_k)=2$, all other $N=5$  topological orders have $N^{ij}_k=0,1$. 
} \label{toplst5} 
\vskip 2mm
\centering
\begin{tabular}{ |c|c|c|c|c| } \hline $N^B_c$ & $S_\text{top}$ &
$d_1,d_2,\cdots$ & $s_1,s_2,\cdots$ & type  \\
\hline 
$5^B_{ 0}$ & $1.1609$ & $1,1,1,1,1$ & $0, \frac{1}{5}, \frac{1}{5},-\frac{1}{5},-\frac{1}{5}$ & 
\\
$5^B_{4}$ & $1.1609$ & $1,1,1,1,1$ & $0, \frac{2}{5}, \frac{2}{5},-\frac{2}{5},-\frac{2}{5}$ & ($A_4,1$)
\\ 
\hline
$5^{B,a}_{ 2}$ & $1.7924$ & $1,1,\zeta_4^1,\zeta_4^1,2$ & $0, 0, \frac{1}{8},-\frac{3}{8}, \frac{1}{3}$ & ($A_1,4$), $(U(1)_3/\Z_2)_{1 \over 2}$
\\
$5^{B,b}_{ 2}$ & $1.7924$ & $1,1,\zeta_4^1,\zeta_4^1,2$ & $0, 0,-\frac{1}{8}, \frac{3}{8}, \frac{1}{3}$ & $ [ 5^{B,a}_{ 2} \otimes 4_0^{B,b}]_{1 \over 4}$
\\
\hline
$5^{B,a}_{-2}$ & $1.7924$ & $1,1,\zeta_4^1,\zeta_4^1,2$ & $0, 0,-\frac{1}{8}, \frac{3}{8},-\frac{1}{3}$ & ($C_4,1$), $(A_3,2)_{1 \over 2}$
\\
$5^{B,b}_{-2}$ & $1.7924$ & $1,1,\zeta_4^1,\zeta_4^1,2$ & $0, 0, \frac{1}{8},-\frac{3}{8},-\frac{1}{3}$ & $ [ 5^{B,a}_{- 2} \otimes 4_0^{B,b}]_{1 \over 4}$
\\
\hline
$5^B_{\frc{16}{11}}$ & $2.5573$ & $1,\zeta_9^1,\zeta_9^2,\zeta_9^3,\zeta_9^4$ & $0,-\frac{2}{11}, \frac{2}{11}, \frac{1}{11},-\frac{5}{11}$ & ($F_4,2$), ($A_1,9)_{1 \over 2}$
\\
$5^B_{-\frc{16}{11}}$ & $2.5573$ & $1,\zeta_9^1,\zeta_9^2,\zeta_9^3,\zeta_9^4$ & $0, \frac{2}{11},-\frac{2}{11},-\frac{1}{11}, \frac{5}{11}$ & ($E_8,3$), $(A_8,2)_{1 \over 9}$
\\ 
$5^B_{\frc{18}{7}}$ & $2.5716$ & $1,\zeta_5^2,\zeta_5^2, \zeta_{12}^2, \zeta_{12}^4$ & $0,-\frac{1}{7},-\frac{1}{7}, \frac{1}{7}, \frac{3}{7}$ & $(A_1,12)_{1 \over 4}$, $(A_2,4)_{1 \over 3}$
\\
$5^B_{-\frc{18}{7}}$ & $2.5716$ & $1,\zeta_5^2,\zeta_5^2, \zeta_{12}^2 , \zeta_{12}^4$ & $0, \frac{1}{7}, \frac{1}{7},-\frac{1}{7},-\frac{3}{7}$ & $(A_3,3)_{1 \over 4}$
\\
\hline 
\end{tabular}
\end{table*}
\def\arraystretch{1.2} \setlength\tabcolsep{3pt} 
\begin{table*}[tb] 
\caption{A list of 50 bosonic rank $N=6$ topological orders in 2+1D with max$(N^{ij}_k)\leq 2$.
} \label{toplst6} 
\vskip 2mm
\centering
\begin{tabular}{ |c|c|c|c|c|c|c| } 
\hline 
$N^B_c$ & $S_\text{top}$ & $D^2$ &
$d_1,d_2,\cdots$ & $s_1,s_2,\cdots$ & $N^B_c\otimes \widetilde{N}^B_{\t c}$ & type \\
 \hline 
 $6^B_{ 1}$ & $1.2924$ & $6$ & $1,1,1,1,1,1$ & $0, \frac{1}{12}, \frac{1}{12},-\frac{1}{4}, \frac{1}{3}, \frac{1}{3}$ & $2^B_{-1}\otimes 3^B_{ 2}$ & $U(1)_3$
 \\
$6^B_{-1}$ & $1.2924$ & $6$ & $1,1,1,1,1,1$ & $0,-\frac{1}{12},-\frac{1}{12}, \frac{1}{4},-\frac{1}{3},-\frac{1}{3}$ & $2^B_{ 1}\otimes 3^B_{-2}$ &
\\
$6^B_{ 3}$ & $1.2924$ & $6$ & $1,1,1,1,1,1$ & $0, \frac{1}{4}, \frac{1}{3}, \frac{1}{3},-\frac{5}{12},-\frac{5}{12}$ & $2^B_{ 1}\otimes 3^B_{ 2}$ &
\\
$6^B_{-3}$ & $1.2924$ & $6$ & $1,1,1,1,1,1$ & $0,-\frac{1}{4},-\frac{1}{3},-\frac{1}{3}, \frac{5}{12}, \frac{5}{12}$ & $2^B_{-1}\otimes 3^B_{-2}$ & ($A_5,1$)
\\
 \hline 
 $6^B_{\frc{1}{2}}$ & $1.5$ & $8$ & $1,1,1,1,\zeta_2^1,\zeta_2^1$ & $0, \frac{1}{4},-\frac{1}{4},\frac{1}{2},-\frac{1}{16}, \frac{3}{16}$ & $2^B_{1}\otimes 3^B_{-\frc{1}{2}}$ &
 \\
$6^B_{-\frc{1}{2}}$ & $1.5$ & $8$ & $1,1,1,1,\zeta_2^1,\zeta_2^1$ & $0, \frac{1}{4},-\frac{1}{4},\frac{1}{2}, \frac{1}{16},-\frac{3}{16}$ & $2^B_{ 1}\otimes 3^B_{-\frc{3}{2}}$ &
\\
$6^B_{\frac{3}{2}}$ & $1.5$ & $8$ & $1,1,1,1,\zeta_2^1,\zeta_2^1$ & $0, \frac{1}{4},-\frac{1}{4},\frac{1}{2}, \frac{1}{16}, \frac{5}{16}$ & $2^B_{ 1}\otimes 3^B_{\frc{1}{2}}$ &
 \\
$6^B_{-\frc{3}{2}}$ & $1.5$ & $8$ & $1,1,1,1,\zeta_2^1,\zeta_2^1$ & $0, \frac{1}{4},-\frac{1}{4},\frac{1}{2},-\frac{1}{16},-\frac{5}{16}$ & $2^B_{1}\otimes 3^B_{-\frc{5}{2}}$ &
\\
$6^B_{\frc{5}{2}}$ & $1.5$ & $8$ & $1,1,1,1,\zeta_2^1,\zeta_2^1$ & $0, \frac{1}{4},-\frac{1}{4},\frac{1}{2}, \frac{3}{16}, \frac{7}{16}$ & $2^B_{ 1}\otimes 3^B_{\frc{3}{2}}$ &
\\
$6^B_{-\frc{5}{2}}$ & $1.5$ & $8$ & $1,1,1,1,\zeta_2^1,\zeta_2^1$ & $0,\frac{1}{4},-\frac{1}{4},\frac{1}{2},-\frac{3}{16},-\frac{7}{16}$ & $2^B_{1}\otimes 3^B_{-\frc{7}{2}}$ &
\\
$6^B_{\frc{7}{2}}$ & $1.5$ & $8$ & $1,1,1,1,\zeta_2^1,\zeta_2^1$ & $0, \frac{1}{4},-\frac{1}{4},\frac{1}{2}, \frac{5}{16},-\frac{7}{16}$ & $2^B_{ 1}\otimes 3^B_{\frc{5}{2}}$ &
\\
$6^B_{-\frc{7}{2}}$ & $1.5$ & $8$ & $1,1,1,1,\zeta_2^1,\zeta_2^1$ & $0,\frac{1}{4},-\frac{1}{4},\frac{1}{2},-\frac{5}{16}, \frac{7}{16}$ & $2^B_{1}\otimes 3^B_{\frc{7}{2}}$ &
\\
\hline 
$6^B_{\frc{4}{5}}$ & $1.7200$ & $10.854$ & $1,1,1,\zeta_3^1,\zeta_3^1,\zeta_3^1$ & $0,-\frac{1}{3},-\frac{1}{3},
\frac{1}{15}, \frac{1}{15}, \frac{2}{5}$ & $2^B_{\frc{14}{5}}\otimes 3^B_{-2}$ &
\\
$6^B_{-\frc{4}{5}}$ & $1.7200$ & $10.854$ & $1,1,1,\zeta_3^1,\zeta_3^1,\zeta_3^1$ & $0, \frac{1}{3},
\frac{1}{3},-\frac{1}{15},-\frac{1}{15},-\frac{2}{5}$ &
$2^B_{-\frc{14}{5}}\otimes 3^B_{ 2}$ &
\\
 $6^B_{\frc{16}{5}}$ & $1.7200$ & $10.854$ &
$1,1,1,\zeta_3^1,\zeta_3^1,\zeta_3^1$ & $0,-\frac{1}{3},-\frac{1}{3},
\frac{4}{15}, \frac{4}{15},-\frac{2}{5}$ & $2^B_{-\frc{14}{5}}\otimes 3^B_{-2}$ & ($A_2,2$)
\\
$6^B_{-\frc{16}{5}}$ & $1.7200$ & $10.854$ &
$1,1,1,\zeta_3^1,\zeta_3^1,\zeta_3^1$ & $0, \frac{1}{3},
\frac{1}{3},-\frac{4}{15},-\frac{4}{15}, \frac{2}{5}$ & $2^B_{
\frac{14}{5}}\otimes 3^B_{ 2}$ &
\\
\hline 
 $6^B_{-\frc{27}{10}}$ & $1.9276$ &
$14.472$ & $1,1,\zeta_2^1,\zeta_3^1,\zeta_3^1,\zeta_2^1\zeta_3^1$ &
$0,\frac{1}{2}, \frac{5}{16},-\frac{1}{10}, \frac{2}{5},-\frac{23}{80}$ & $2^B_{
\frac{14}{5}}\otimes 3^B_{\frc{5}{2}}$ & ($E_7,2$)
\\
 $6^B_{-\frc{17}{10}}$ & $1.9276$ &
$14.472$ & $1,1,\zeta_2^1,\zeta_3^1,\zeta_3^1,\zeta_2^1\zeta_3^1$ &
$0,\frac{1}{2}, \frac{7}{16},-\frac{1}{10}, \frac{2}{5},-\frac{13}{80}$ & $2^B_{
\frac{14}{5}}\otimes 3^B_{\frc{7}{2}}$ & 
\\
 $6^B_{-\frc{7}{10}}$ & $1.9276$ &
$14.472$ & $1,1,\zeta_2^1,\zeta_3^1,\zeta_3^1,\zeta_2^1\zeta_3^1$ &
$0,\frac{1}{2},-\frac{7}{16},-\frac{1}{10}, \frac{2}{5},-\frac{3}{80}$ & $2^B_{
\frac{14}{5}}\otimes 3^B_{-\frc{7}{2}}$ & 
\\
$6^B_{\frc{3}{10}}$ & $1.9276$ & $14.472$
& $1,1,\zeta_2^1,\zeta_3^1,\zeta_3^1,\zeta_2^1\zeta_3^1$ &
$0,\frac{1}{2},-\frac{5}{16},-\frac{1}{10}, \frac{2}{5}, \frac{7}{80}$ & $2^B_{
\frac{14}{5}}\otimes 3^B_{-\frc{5}{2}}$ & 
\\
 $6^B_{\frc{13}{10}}$ & $1.9276$ &
$14.472$ & $1,1,\zeta_2^1,\zeta_3^1,\zeta_3^1,\zeta_2^1\zeta_3^1$ &
$0,\frac{1}{2},-\frac{3}{16},-\frac{1}{10}, \frac{2}{5}, \frac{17}{80}$ & $2^B_{
\frac{14}{5}}\otimes 3^B_{-\frc{3}{2}}$ & 
\\
 $6^B_{\frc{23}{10}}$ & $1.9276$ &
$14.472$ & $1,1,\zeta_2^1,\zeta_3^1,\zeta_3^1,\zeta_2^1\zeta_3^1$ &
$0,\frac{1}{2},-\frac{1}{16},-\frac{1}{10}, \frac{2}{5}, \frac{27}{80}$ & $2^B_{
\frac{14}{5}}\otimes 3^B_{-\frc{1}{2}}$ & 
\\
 $6^B_{\frc{33}{10}}$ & $1.9276$ &
$14.472$ & $1,1,\zeta_2^1,\zeta_3^1,\zeta_3^1,\zeta_2^1\zeta_3^1$ &
$0,\frac{1}{2}, \frac{1}{16},-\frac{1}{10}, \frac{2}{5}, \frac{37}{80}$ & $2^B_{
\frac{14}{5}}\otimes 3^B_{\frc{1}{2}}$ & 
\\
 $6^B_{-\frc{37}{10}}$ & $1.9276$ &
$14.472$ & $1,1,\zeta_2^1,\zeta_3^1,\zeta_3^1,\zeta_2^1\zeta_3^1$ &
$0,\frac{1}{2}, \frac{3}{16},-\frac{1}{10}, \frac{2}{5},-\frac{33}{80}$ & $2^B_{
\frac{14}{5}}\otimes 3^B_{\frc{3}{2}}$ & 
\\
\hline
 $6^B_{\frc{27}{10}}$ & $1.9276$ &
$14.472$ & $1,1,\zeta_2^1,\zeta_3^1,\zeta_3^1,\zeta_2^1\zeta_3^1$ &
$0,\frac{1}{2},-\frac{5}{16}, \frac{1}{10},-\frac{2}{5}, \frac{23}{80}$ &
$2^B_{-\frc{14}{5}}\otimes 3^B_{-\frc{5}{2}}$ & 
\\
 $6^B_{\frc{17}{10}}$ & $1.9276$ &
$14.472$ & $1,1,\zeta_2^1,\zeta_3^1,\zeta_3^1,\zeta_2^1\zeta_3^1$ &
$0,\frac{1}{2},-\frac{7}{16}, \frac{1}{10},-\frac{2}{5}, \frac{13}{80}$ &
$2^B_{-\frc{14}{5}}\otimes 3^B_{-\frc{7}{2}}$ & 
\\
 $6^B_{\frc{7}{10}}$ & $1.9276$ &
$14.472$ & $1,1,\zeta_2^1,\zeta_3^1,\zeta_3^1,\zeta_2^1\zeta_3^1$ &
$0,\frac{1}{2}, \frac{7}{16}, \frac{1}{10},-\frac{2}{5}, \frac{3}{80}$ &
$2^B_{-\frc{14}{5}}\otimes 3^B_{\frc{7}{2}}$ & 
\\
 $6^B_{-\frc{3}{10}}$ & $1.9276$ &
$14.472$ & $1,1,\zeta_2^1,\zeta_3^1,\zeta_3^1,\zeta_2^1\zeta_3^1$ &
$0,\frac{1}{2}, \frac{5}{16}, \frac{1}{10},-\frac{2}{5},-\frac{7}{80}$ &
$2^B_{-\frc{14}{5}}\otimes 3^B_{\frc{5}{2}}$ & 
\\
 $6^B_{-\frc{13}{10}}$ & $1.9276$ &
$14.472$ & $1,1,\zeta_2^1,\zeta_3^1,\zeta_3^1,\zeta_2^1\zeta_3^1$ &
$0,\frac{1}{2}, \frac{3}{16}, \frac{1}{10},-\frac{2}{5},-\frac{17}{80}$ &
$2^B_{-\frc{14}{5}}\otimes 3^B_{\frc{3}{2}}$ & 
\\
 $6^B_{-\frc{23}{10}}$ & $1.9276$ &
$14.472$ & $1,1,\zeta_2^1,\zeta_3^1,\zeta_3^1,\zeta_2^1\zeta_3^1$ &
$0,\frac{1}{2}, \frac{1}{16}, \frac{1}{10},-\frac{2}{5},-\frac{27}{80}$ &
$2^B_{-\frc{14}{5}}\otimes 3^B_{\frc{1}{2}}$ & 
\\
 $6^B_{-\frc{33}{10}}$ & $1.9276$ &
$14.472$ & $1,1,\zeta_2^1,\zeta_3^1,\zeta_3^1,\zeta_2^1\zeta_3^1$ &
$0,\frac{1}{2},-\frac{1}{16}, \frac{1}{10},-\frac{2}{5},-\frac{37}{80}$ &
$2^B_{-\frc{14}{5}}\otimes 3^B_{-\frc{1}{2}}$ & 
\\
 $6^B_{\frc{37}{10}}$ & $1.9276$ &
$14.472$ & $1,1,\zeta_2^1,\zeta_3^1,\zeta_3^1,\zeta_2^1\zeta_3^1$ &
$0,\frac{1}{2},-\frac{3}{16}, \frac{1}{10},-\frac{2}{5}, \frac{33}{80}$ &
$2^B_{-\frc{14}{5}}\otimes 3^B_{-\frc{3}{2}}$ & 
\\
 \hline 
 $6^B_{\frc{1}{7}}$ & $2.1082$ &
$18.591$ & $1,1,\zeta_5^1,\zeta_5^1,\zeta_5^2,\zeta_5^2$ &
$0,-\frac{1}{4},-\frac{1}{7},-\frac{11}{28}, \frac{1}{28}, \frac{2}{7}$ &
$2^B_{-1}\otimes 3^B_{\frc{8}{7}}$ &
\\
 $6^B_{-\frc{1}{7}}$ & $2.1082$ & $18.591$ &
$1,1,\zeta_5^1,\zeta_5^1,\zeta_5^2,\zeta_5^2$ & $0, \frac{1}{4}, \frac{1}{7},
\frac{11}{28},-\frac{1}{28},-\frac{2}{7}$ & $2^B_{ 1}\otimes 3^B_{-\frc{8}{7}}$ & ($C_5,1$)
\\
$6^B_{\frc{15}{7}}$ & $2.1082$ & $18.591$ &
$1,1,\zeta_5^1,\zeta_5^1,\zeta_5^2,\zeta_5^2$ & $0, \frac{1}{4},
\frac{3}{28},-\frac{1}{7}, \frac{2}{7},-\frac{13}{28}$ & $2^B_{ 1}\otimes 3^B_{
\frac{8}{7}}$ & ($A_1,5$)
\\
 $6^B_{-\frc{15}{7}}$ & $2.1082$ & $18.591$ &
$1,1,\zeta_5^1,\zeta_5^1,\zeta_5^2,\zeta_5^2$ & $0,-\frac{1}{4},-\frac{3}{28},
\frac{1}{7},-\frac{2}{7}, \frac{13}{28}$ & $2^B_{-1}\otimes 3^B_{-\frc{8}{7}}$ &
\\
\hline $6^{B,a}_{ 0}$ & $2.1609$ & $20$ & $1,1,2,2,\sqrt{5},\sqrt{5}$ & $0, 0,
\frac{1}{5},-\frac{1}{5}, 0,\frac{1}{2}$ & & $(D_5,2)_{1 \over 2}$, $(U(1)_5/\Z_2)_{1 \over 2}$
\\
 $6^{B,b}_{ 0}$ & $2.1609$ & $20$ &
$1,1,2,2,\sqrt{5},\sqrt{5}$ & $0, 0, \frac{1}{5},-\frac{1}{5},
\frac{1}{4},-\frac{1}{4}$ & &  $[6^{B,a}_{0}\otimes 4^{B,b}_0]_{\frac14}$ 
\\
\hline
 $6^{B,b}_{4}$ & $2.1609$ & $20$ & $1,1,2,2,\sqrt{5},\sqrt{5}$ & $0, 0,
\frac{2}{5},-\frac{2}{5}, \frac{1}{4},-\frac{1}{4}$ & & ($B_2,2$)
\\
 $6^{B,a}_{4}$ & $2.1609$ & $20$ &
$1,1,2,2,\sqrt{5},\sqrt{5}$ & $0, 0, \frac{2}{5},-\frac{2}{5}, 0,\frac{1}{2}$ & & $[6^{B,b}_{4}\otimes 4^{B,b}_0]_{\frac14}$
\\
 \hline $6^B_{
\frc{58}{35}}$ & $2.5359$ & $33.632$ &
$1,\zeta_3^1,\zeta_5^1,\zeta_5^2,\zeta_3^1\zeta_5^1,\zeta_3^1\zeta_5^2$ & $0,
\frac{2}{5}, \frac{1}{7},-\frac{2}{7},-\frac{16}{35}, \frac{4}{35}$ & $2^B_{
\frac{14}{5}}\otimes 3^B_{-\frc{8}{7}}$ &
\\
 $6^B_{-\frc{58}{35}}$ & $2.5359$ &
$33.632$ &
$1,\zeta_3^1,\zeta_5^1,\zeta_5^2,\zeta_3^1\zeta_5^1,\zeta_3^1\zeta_5^2$ &
$0,-\frac{2}{5},-\frac{1}{7}, \frac{2}{7}, \frac{16}{35},-\frac{4}{35}$ &
$2^B_{-\frc{14}{5}}\otimes 3^B_{\frc{8}{7}}$ &
\\
 $6^B_{\frc{138}{35}}$ & $2.5359$
& $33.632$ &
$1,\zeta_3^1,\zeta_5^1,\zeta_5^2,\zeta_3^1\zeta_5^1,\zeta_3^1\zeta_5^2$ & $0,
\frac{2}{5},-\frac{1}{7}, \frac{2}{7}, \frac{9}{35},-\frac{11}{35}$ & $2^B_{
\frac{14}{5}}\otimes 3^B_{\frc{8}{7}}$ &
\\
 $6^B_{-\frc{138}{35}}$ & $2.5359$ &
$33.632$ &
$1,\zeta_3^1,\zeta_5^1,\zeta_5^2,\zeta_3^1\zeta_5^1,\zeta_3^1\zeta_5^2$ &
$0,-\frac{2}{5}, \frac{1}{7},-\frac{2}{7},-\frac{9}{35}, \frac{11}{35}$ &
$2^B_{-\frc{14}{5}}\otimes 3^B_{-\frc{8}{7}}$ &
\\
 \hline {$6^B_{\frc{46}{13}}$}
& {$2.9132$} & {$56.746$} &
{$1,\zeta_{11}^{1},\zeta_{11}^{2},\zeta_{11}^{3},\zeta_{11}^{4},\zeta_{11}^{5}$}
& {$0, \frac{4}{13}, \frac{2}{13},-\frac{6}{13},
\frac{6}{13},-\frac{1}{13}$} & & ($A_1,11)_{1 \over 2}$
\\
 {$6^B_{-\frc{46}{13}}$} & {$2.9132$} &
{$56.746$} &
{$1,\zeta_{11}^{1},\zeta_{11}^{2},\zeta_{11}^{3},\zeta_{11}^{4},\zeta_{11}^{5}$}
& {$0,-\frac{4}{13},-\frac{2}{13}, \frac{6}{13},-\frac{6}{13},
\frac{1}{13}$} & & $(A_{10},2)_{1 \over 11}$
\\
 \hline $6^B_{\frc{8}{3}}$ & $3.1107$ &
$74.617$ &
$1,\zeta_{7}^{3},\zeta_{7}^{3},\zeta_{16}^{2},\zeta_{16}^{4},\zeta_{16}^{6}$
& $0, \frac{1}{9}, \frac{1}{9}, \frac{1}{9}, \frac{1}{3},-\frac{1}{3}$ & & $(A_1,16)_{1 \over 4}$
\\
$6^B_{-\frc{8}{3}}$ & $3.1107$ & $74.617$ &
$1,\zeta_{7}^{3},\zeta_{7}^{3},\zeta_{16}^{2},\zeta_{16}^{4},\zeta_{16}^{6}$
& $0,-\frac{1}{9},-\frac{1}{9},-\frac{1}{9},-\frac{1}{3}, \frac{1}{3}$ & & $(A_2,6)_{1 \over 9}$
\\
 \hline 
$6^B_{2}$ & $3.3263$ & $100.61$ & $1,\frac{3+\sqrt{21}}{2},\frac{3+\sqrt{21}}{2},\frac{3+\sqrt{21}}{2},\frac{5+\sqrt{21}}{2},\frac{7+\sqrt{21}}{2}$ & $0,-\frac{1}{7},-\frac{2}{7}, \frac{3}{7}, 0, \frac{1}{3}$ & & \\
$6^B_{-2}$ & $3.3263$ & $100.61$ & $1,\frac{3+\sqrt{21}}{2},\frac{3+\sqrt{21}}{2},\frac{3+\sqrt{21}}{2},\frac{5+\sqrt{21}}{2},\frac{7+\sqrt{21}}{2}$ & $0, \frac{1}{7}, \frac{2}{7},-\frac{3}{7}, 0,-\frac{1}{3}$ & & $(G_2,3)$ \\
 \hline 
\end{tabular} 
\end{table*}

\begin{table*}[tb] 
\caption{ A list of 24 bosonic rank $N=7$ topological
orders in 2+1D with max$(N^{ij}_k)\leq 1$.   Since $N=7$ is a prime number, all
those 24 topological orders are primitive.  
} \label{toplst7} 
\vskip 2mm
\centering
\begin{tabular}{ |c|c|c|c|c|c| } \hline $N^B_c$ & $S_\text{top}$ & $D^2$ &
$d_1,d_2,\cdots$ & $s_1,s_2,\cdots$  & type \\
 \hline 
 $7^{B,a}_{ 2}$ & $1.4036$ & $7$ &
$1,1,1,1,1,1,1$ & $0, \frac{1}{7}, \frac{1}{7}, \frac{2}{7},
\frac{2}{7},-\frac{3}{7},-\frac{3}{7}$ &
\\
 $7^{B,a}_{-2}$ & $1.4036$ & $7$ &
$1,1,1,1,1,1,1$ & $0,-\frac{1}{7},-\frac{1}{7},-\frac{2}{7},-\frac{2}{7},
\frac{3}{7}, \frac{3}{7}$ & ($A_6,1$)
\\
 \hline 
 $7^B_{\frc{9}{4}}$ & $2.3857$ &
$27.313$ & $1,1,\zeta_6^1,\zeta_6^1,\zeta_6^2,\zeta_6^2,\zeta_6^3$ &
$0,\frac{1}{2}, \frac{3}{32}, \frac{3}{32}, \frac{1}{4},-\frac{1}{4},
\frac{15}{32}$ & ($A_1,6$)
\\
 $7^B_{\frc{13}{4}}$ & $2.3857$ & $27.313$ &
$1,1,\zeta_6^1,\zeta_6^1,\zeta_6^2,\zeta_6^2,\zeta_6^3$ & $0,\frac{1}{2},
\frac{7}{32}, \frac{7}{32}, \frac{1}{4},-\frac{1}{4},-\frac{13}{32}$ & $[7^B_{\frc{9}{4}}\otimes 4^B_1]_{\frac14}$
\\
 $7^B_{-\frc{15}{4}}$ & $2.3857$ & $27.313$ &
$1,1,\zeta_6^1,\zeta_6^1,\zeta_6^2,\zeta_6^2,\zeta_6^3$ & $0,\frac{1}{2},
\frac{11}{32}, \frac{11}{32}, \frac{1}{4},-\frac{1}{4},-\frac{9}{32}$ & $[7^B_{\frc{13}{4}}\otimes 4^B_1]_{\frac14}$ 
\\
 $7^B_{-\frc{11}{4}}$ & $2.3857$ & $27.313$ &
$1,1,\zeta_6^1,\zeta_6^1,\zeta_6^2,\zeta_6^2,\zeta_6^3$ & $0,\frac{1}{2},
\frac{15}{32}, \frac{15}{32}, \frac{1}{4},-\frac{1}{4},-\frac{5}{32}$ & $[7^B_{-\frc{15}{4}}\otimes 4^B_1]_{\frac14}$
\\
$7^B_{-\frc{7}{4}}$ & $2.3857$ & $27.313$ &
$1,1,\zeta_6^1,\zeta_6^1,\zeta_6^2,\zeta_6^2,\zeta_6^3$ &
$0,\frac{1}{2},-\frac{13}{32},-\frac{13}{32},
\frac{1}{4},-\frac{1}{4},-\frac{1}{32}$ & $[7^B_{-\frc{11}{4}}\otimes 4^B_1]_{\frac14}$
\\
$7^B_{-\frc{3}{4}}$ & $2.3857$ & $27.313$ &
$1,1,\zeta_6^1,\zeta_6^1,\zeta_6^2,\zeta_6^2,\zeta_6^3$ &
$0,\frac{1}{2},-\frac{9}{32},-\frac{9}{32}, \frac{1}{4},-\frac{1}{4},
\frac{3}{32}$ & $[7^B_{-\frc{7}{4}}\otimes 4^B_1]_{\frac14}$
\\
$7^B_{\frc{1}{4}}$ & $2.3857$ & $27.313$ &
$1,1,\zeta_6^1,\zeta_6^1,\zeta_6^2,\zeta_6^2,\zeta_6^3$ &
$0,\frac{1}{2},-\frac{5}{32},-\frac{5}{32}, \frac{1}{4},-\frac{1}{4},
\frac{7}{32}$ & $[7^B_{-\frc{3}{4}}\otimes 4^B_1]_{\frac14}$
 \\
 $7^B_{\frc{5}{4}}$ & $2.3857$ & $27.313$ &
$1,1,\zeta_6^1,\zeta_6^1,\zeta_6^2,\zeta_6^2,\zeta_6^3$ &
$0,\frac{1}{2},-\frac{1}{32},-\frac{1}{32}, \frac{1}{4},-\frac{1}{4},
\frac{11}{32}$ & $[7^B_{\frc{1}{4}}\otimes 4^B_1]_{\frac14}$
\\
\hline
 $7^B_{\frc{7}{4}}$ & $2.3857$ & $27.313$ &
$1,1,\zeta_6^1,\zeta_6^1,\zeta_6^2,\zeta_6^2,\zeta_6^3$ & $0,\frac{1}{2},
\frac{13}{32}, \frac{13}{32}, \frac{1}{4},-\frac{1}{4}, \frac{1}{32}$ & ($C_6,1$)
\\
 $7^B_{\frc{11}{4}}$ & $2.3857$ &
$27.313$ & $1,1,\zeta_6^1,\zeta_6^1,\zeta_6^2,\zeta_6^2,\zeta_6^3$ &
$0,\frac{1}{2},-\frac{15}{32},-\frac{15}{32}, \frac{1}{4},-\frac{1}{4},
\frac{5}{32}$ & $[7^B_{\frc{7}{4}}\otimes 4^B_1]_{\frac14}$
\\
 $7^B_{\frc{15}{4}}$ & $2.3857$ & $27.313$ &
$1,1,\zeta_6^1,\zeta_6^1,\zeta_6^2,\zeta_6^2,\zeta_6^3$ &
$0,\frac{1}{2},-\frac{11}{32},-\frac{11}{32}, \frac{1}{4},-\frac{1}{4},
\frac{9}{32}$ & $[7^B_{\frc{11}{4}}\otimes 4^B_1]_{\frac14}$
\\
$7^B_{-\frc{13}{4}}$ & $2.3857$ & $27.313$ &
$1,1,\zeta_6^1,\zeta_6^1,\zeta_6^2,\zeta_6^2,\zeta_6^3$ &
$0,\frac{1}{2},-\frac{7}{32},-\frac{7}{32}, \frac{1}{4},-\frac{1}{4},
\frac{13}{32}$ & $[7^B_{\frc{15}{4}}\otimes 4^B_1]_{\frac14}$
\\
 $7^B_{-\frc{9}{4}}$ & $2.3857$ & $27.313$ &
$1,1,\zeta_6^1,\zeta_6^1,\zeta_6^2,\zeta_6^2,\zeta_6^3$ &
$0,\frac{1}{2},-\frac{3}{32},-\frac{3}{32},
\frac{1}{4},-\frac{1}{4},-\frac{15}{32}$ & $[7^B_{-\frc{13}{4}}\otimes 4^B_1]_{\frac14}$
\\
 $7^B_{-\frc{5}{4}}$ & $2.3857$ & $27.313$ &
$1,1,\zeta_6^1,\zeta_6^1,\zeta_6^2,\zeta_6^2,\zeta_6^3$ & $0,\frac{1}{2},
\frac{1}{32}, \frac{1}{32}, \frac{1}{4},-\frac{1}{4},-\frac{11}{32}$ & $[7^B_{-\frc{9}{4}}\otimes 4^B_1]_{\frac14}$
\\
 $7^B_{-\frc{1}{4}}$ & $2.3857$ & $27.313$ &
$1,1,\zeta_6^1,\zeta_6^1,\zeta_6^2,\zeta_6^2,\zeta_6^3$ & $0,\frac{1}{2},
\frac{5}{32}, \frac{5}{32}, \frac{1}{4},-\frac{1}{4},-\frac{7}{32}$ &  
$[7^B_{-\frc{5}{4}}\otimes 4^B_1]_{\frac14}$
\\
 $7^B_{
\frc{3}{4}}$ & $2.3857$ & $27.313$ &
$1,1,\zeta_6^1,\zeta_6^1,\zeta_6^2,\zeta_6^2,\zeta_6^3$ & $0,\frac{1}{2},
\frac{9}{32}, \frac{9}{32}, \frac{1}{4},-\frac{1}{4},-\frac{3}{32}$ & $[7^B_{-\frc{1}{4}}\otimes 4^B_1]_{\frac14}$
\\
\hline
$7^{B,b}_{ 2}$ & $2.4036$ & $28$ & $1,1,2,2,2,\sqrt{7},\sqrt{7}$ & $0, 0,
\frac{1}{7}, \frac{2}{7},-\frac{3}{7}, \frac{1}{8},-\frac{3}{8}$ & $(U(1)_7/\Z_2)_{1 \over 2}$
\\
 $7^{B,c}_{ 2}$ & $2.4036$ & $28$ & $1,1,2,2,2,\sqrt{7},\sqrt{7}$ & $0, 0, \frac{1}{7},
\frac{2}{7},-\frac{3}{7},-\frac{1}{8}, \frac{3}{8}$ & $[7^{B,b}_{2} \otimes 4_0^{B,b}]_{1 \over 4}$
\\
\hline
 $7^{B,b}_{-2}$ & $2.4036$ & $28$ &
$1,1,2,2,2,\sqrt{7},\sqrt{7}$ & $0, 0,-\frac{1}{7},-\frac{2}{7},
\frac{3}{7},-\frac{1}{8}, \frac{3}{8}$ & $(B_3,2)$, $(D_7,2)_{1 \over 2}$
\\
 $7^{B,c}_{-2}$ & $2.4036$ &
$28$ & $1,1,2,2,2,\sqrt{7},\sqrt{7}$ & $0, 0,-\frac{1}{7},-\frac{2}{7},
\frac{3}{7}, \frac{1}{8},-\frac{3}{8}$ & $[7^{B,b}_{-2} \otimes 4_0^{B,b}]_{1 \over 4}$
\\
\hline 
$7^B_{8/5}$ & $3.2194$ & $86.750$ & $1,\zeta_{13}^{1},\zeta_{13}^{2},\zeta_{13}^{3},\zeta_{13}^{4},\zeta_{13}^{5},\zeta_{13}^{6}$ & $0,-\frac{1}{5}, \frac{2}{15}, 0, \frac{2}{5}, \frac{1}{3},-\frac{1}{5}$ & $(A_1,13)_{1 \over 2}$ \\
$7^B_{-8/5}$ & $3.2194$ & $86.750$ & $1,\zeta_{13}^{1},\zeta_{13}^{2},\zeta_{13}^{3},\zeta_{13}^{4},\zeta_{13}^{5},\zeta_{13}^{6}$ & $0, \frac{1}{5},-\frac{2}{15}, 0,-\frac{2}{5},-\frac{1}{3}, \frac{1}{5}$ & $(A_{12},2)_{1 \over 13}$\\
\hline 
\end{tabular} 
\end{table*}

If we further require that
\begin{align}
\sum_{IJ}  a^I K_{IJ}  a^J 
 -   2 \hsc{\vec a} +\hsc{2\vec a} = \text{ even},
\end{align}
for all $\vec a$,
then we will obtain the bosonic 2+1D topological orders of a
given non-abelian type. 
If we require that
\begin{align}
\sum_{IJ}  a^I K_{IJ} a^J 
 -   2 \hsc{\vec a} +\hsc{2\vec a} = \text{ odd},
\end{align}
for some $\vec a$,
then we will obtain the fermionic 2+1D topological orders of a
given non-abelian type. 

\subsection{Topological excitations from simple-current algebra}

In the above, we have used the simple-current CFT generated by the simple
currents $c_I(z)$ to obtain the ground state wave function of a 2+1D
topological order.  In this section, we are going to discuss how to obtain the
topological excitations from the simple-current CFT.

First, we like to introduce the notion of simple-current primary field.  Acting
with $c_I(z)$ on to the ground state $|0\>$ generates the adjoint representation of the
simple-current algebra.  The simple-current algebra has other irreducible
representations, which can be obtained by the action of $c_I(z)$ on the ground state
$|\eta\>=\eta|0\>$ of a twisted sector. Thus the different irreducible
representations of the simple-current algebra are labeled by $\eta$ (where
$\eta=1$ corresponds the adjoint representation).  The operator $\eta(z)$ that
corresponds to the twisted ground state $|\eta\>$ under the operator-state
correspondence is called a primary field of the simple-current algebra.

The primary fields $\eta(z)$ are local with respect to all the simple currents
$c_I(z)$:
\begin{align}
 c_I(z)\eta(w) \sim (z-w)^{\al_{c_I,\eta}} [c_I\eta](w) + \ldots
\end{align}
where $\al_{c_I,\eta}$ are integers. Each simple-current primary field (or each
irreducible representation of the  simple-current algebra) corresponds to a type
of topological excitation in the corresponding topological order.

So to use CFT to study 2+1D topological order, we need to first identify the
simple currents to produce the many-body wavefunction of the topological order.
We then need to find the irreducible representations (or the primary fields) of
the simple-current algebra to obtain the topological excitations and their
properties (such as the quantum dimensions, the spins, \etc).

In general, the simple currents $c_I(z)$ have the form
\begin{align}
 c_I(z) = \psi_I \ee^{\ii \v k^I \cdot \v \phi}
 =\psi_{I} \ee^{\ii \sum_\mu  k^I_\mu \phi^\mu}
\end{align}
where $\psi_I$ are simple currents with finite order (see \eqn{psin}).
Let us introduce
\begin{align}
\psi_{\vec b}\equiv \prod_I \psi_I^{b_I},\ \ \ \ \
c_{\vec b}\equiv \prod_I c_I^{b_I} .
\end{align}
Also, let us use $\si_\al$, $\al=1,2,\cdots$, to denote the primary fields of
the simple-current CFT generated by simple currents $\psi_I$, and  use
$\si_{\al;\vec b}$ to denote the product of $\si_\al$ and  
$\psi_{\vec b}$. $\si_{\al;\vec b}$ are descendent fields
of the primary field $\si_\al$ and have higher conformal dimensions
\begin{align}
 h^\text{sc}_{\al,\vec b} \geq h^\text{sc}_{\al},
\end{align}
where
 $h^\text{sc}_{\al}$ is the conformal dimension of $\si_{\al}$
and
 $h^\text{sc}_{\al,\vec b}$ is the conformal dimension of $\si_{\al;\vec b}$.
The OPE of $\si_{\al;\vec b}$ with $\psi_{\vec a}$ has for its leading term
\begin{align}
 \psi_{\vec a} (z) \si_{\al;\vec b} (w) \sim 
\frac{1}{(z-w)^{ h^\text{sc}_{\vec a}  +h^\text{sc}_{\al,\vec b}-  
h^\text{sc}_{\al,\vec a+\vec b}}  
}\si_{\al;\vec a+\vec b} .
\end{align}

The simple-current primary field $\eta$ for the original
simple currents $c_I$ is given by
\begin{align}
 \eta_{\al,\v l^\al}=
 \si_{\al} \ee^{\ii \v l^\al\cdot \v \phi}
 =\si_{\al} \ee^{\ii \sum_\mu  l^\al_\mu \phi^\mu} \ .
\end{align} 
The corresponding descendent fields are given by
\begin{align}
 \eta_{\al,\v l^\al;\vec b}=
 \si_{\al,\vec b} \ee^{\ii \v l^\al\cdot \v \phi} \ee^{\ii \sum_I b_I 
\v k^I\cdot \v \phi}
\end{align} 
for all different integer vectors $\vec b$.
Each of those operators should be mutually local with respect to $c_{\vec a}$.
This requires $\v l^\al$ to satisfy
\begin{align}
\label{qpcnd}
\sum_{IJ}  a^I K_{IJ}b^J 
+\sum_{IJ\mu\nu}  a^I k^I_\mu G^{\mu\nu}l^\al_\nu 
-   \hsc{\vec a} -\hsc{\al,\vec b} +\hsc{\al,\vec a+\vec b} \in \Z ,
\end{align}
for any integer vectors $\vec a$ and $\vec b$.

To understand the above construction in more detail, let us count the number of
$c_I$-simple-current primary fields $\eta$ (which is equal to the number of
topological types of the topological excitations in the corresponding
topological order).  First a $c_I$-primary field $\eta$ corresponds to a pair: a
$\psi_I$-simple-current primary field $\si_\al$ and a vector $\v l^\al$.  So
the $c_I$-primary fields are labeled by $(\al,\v l^\al)$.  We have used
$\eta_{\al,\v l^\al}$ to denote those $c_I$-primary fields.  $\v l^\al$ must
satisfy \eqn{qpcnd}.
In fact, it is enough to find rational vectors $\v l^\al$ that satisfy
\begin{align}
\label{Klh}
\sum_{IJ\mu\nu}  a^I k^I_\mu G^{\mu\nu}l^\al_\nu 
-\hsc{\vec a} -\hsc{\al} +\hsc{\al,\vec a} \in \Z, \ \ \ \forall \vec a .
\end{align}
For each $\al$, we may have 
many solutions $\v l^\al$ which satisfy the above equation.
But two solutions $\v l^\al_1$ and $\v l^\al_2$
are regarded as the same if they are related by
\begin{align}
\v l^\al_1-\v l^\al_2
 = \sum_{IJ} \v k^I n_{IJ} a_J , \ \ \ a_J\in \Z. 
\end{align}
Counting the pairs $(\si_\al, \v l^\al)$ of inequivalent solutions will
give us the number of $c_I$-simple-current primary fields  and the  number
of topological types.

\section{2+1D topological order from chiral $\frac{U(1)_M}{\Z_2}$ orbifold CFT}
\label{orb}

In this section, we will give an example of using simple-current algebra to
construct a wavefunction that realizes a 2+1D topological order.  In the
process, we will give a brief review on the $\frac{U(1)_M}{\Z_2}$
orbifold CFT, following \Ref{DVV8985}.  

\subsection{Virasoro algebra}

Here, we will view a CFT as a 1+1D gapless system with unit velocity $v=1$ on a
1D ring of size $2\pi$.  The total Hilbert space $\cV$ of the CFT can always be
viewed as a sum of (irreducible) representations of the Virasoro algebra.  The
Virasoro algebra is generated by the energy-momentum tensor $T(z)$, whose
Fourier components $T(z)=\sum_n z^{-n-2} L_n$ satisfy
\begin{align}
 [L_m, L_n]= (m-n) L_{m+n} +\frac{c}{12}(m^3-m) \del_{m+n,0}.
\end{align}
The character of a representation of the Virasoro algebra is defined as
\begin{align}
 \chi^\text{Vir}_{c,h} (q) =\Tr(q^{L_0-{c \over 24}}) ,
\end{align}
where $L_0=H$ is the Hamiltonian of the CFT on the ring.  So the character
encodes the energy spectrum of the CFT on the ring.  The irreducible
representations of the Virasoro algebra are labeled by $(c,h)$, where $h$ is
the energy of the lowest energy state in the representation. $h$ is also the
conformal dimension of the Virasoro primary field associated to the
representation.  The character of the corresponding irreducible representation
has the general form
\begin{align}
\chi_{c,h}^\text{Vir}(q) =
\frac{q^{1-c \over 24}}{\eta(q)} [q^h -  q^{h_1} + q^{h_2} - \ldots ],
\end{align}
where
\begin{align}
\eta(q)=q^{1 \over 24} \prod_{n=1}^\infty (1-q^n)
\end{align}
and the terms $-q^{h_1} + q^{h_2} + \ldots$ represent subtractions due to null states in the 
Verma module with highest weight $h$.

\subsection{ $U(1)$ current algebra}

The $U(1)$ current algebra (which is a simple-current algebra) is generated by
$j=\ii \prt \phi$. ($j$ is a simple current.)  In other words, the space
$\cV^{U(1)}_1$ of the adjoint representation of the $U(1)$ current algebra is
generated by $j(z)$ acting on the ground state $|0\>$.  The corresponding
primary field for the adjoint representation is the identity operator $1$.
This is why we use $\cV^{U(1)}_1$ to denote the adjoint representation.  The
adjoint representation $\cV^{U(1)}_1$ has a character
$\chi^{U(1)}_1(q)=1/\eta(q)$.  However, the adjoint irreducible representation
of the $U(1)$ current algebra is not an irreducible representation of the
Virasoro algebra. Instead, it is formed by many irreducible representations
$\cV_{c=1,n^2}^\text{Vir}$ of the Virasoro algebra generated by the energy
momentum tensor $T(z)\propto j^2(z)$.  It turns out that
\begin{align}
 \cV^{U(1)}_1 = \bigoplus_{n\geq 0} \cV^\text{Vir}_{c=1,n^2}
\end{align}
since
\begin{align}
 \chi^{U(1)}_1(q)=\frac{1}{\eta(q)}
=\sum_{n\geq 0} \frac{q^{n^2}-q^{(n+1)^2}}{\eta(q)}
=\sum_{n\geq 0} \chi^\text{Vir}_{c=1, n^2} .
\end{align}
The corresponding Virasoro primary fields are 
$1,j,j_4=j^4-2j\prt^2 j +\frac32 (\prt j)^2,\cdots$.

The non-trivial representation $\cV^{U(1)}_k$ of the $U(1)$ current algebra
corresponds to the $U(1)$ primary field $\ee^{\ii k\phi}$ with conformal
dimension $h=\frac{k^2}{2}$.  The corresponding representations have the
following fusion property
\begin{align}
\cV^{U(1)}_k \otimes \cV^{U(1)}_{k'} = \cV^{U(1)}_{k+k'}, \ \
k,k' \in \R . 
\end{align}

\subsection{ Extended $U(1)_M$ current algebra}

The extended $U(1)_M$ current algebra $\cV^{U(1)_M}_1$ of level $M$ (which is
another simple-current algebra) is generated by the spin-$M$ fields $\psi_+=\ee^{\ii
\sqrt{2M}\phi}$ and $\psi_-=\ee^{-\ii \sqrt{2M}\phi}$.  Note that the OPE
$\psi_+\psi_- \sim 1+j$.  So the  extended $U(1)_M$ current algebra is also
generated by $j,\psi_+,\psi_-$.

The non-trivial representation $\cV^{U(1)_M}_k$ of the extended $U(1)_M$
current algebra corresponds to the extended-$U(1)_M$ primary fields $\ee^{\ii
k\phi/\sqrt{2M}}$, $k=0,\cdots,2M-1$, which are local with respect to the 
generating fields $\psi_\pm$.  The corresponding character is given by
\begin{align}
 \chi^{U(1)_M}_k(q)=\frac{1}{\eta(q)}\sum_{m\in \Z} q^{(k+2m M)^2/M} .
\end{align}
Under the modular transformation $S$, the characters $\chi^{U(1)_M}_k(q)$
transforms as
\begin{align}
 S: \chi^{U(1)_M}_k(q) \to \sum_{k'\in \Z_{2M}}
\ee^{-\ii \pi kk'/2M} \chi^{U(1)_M}_{k'}(q) .
\end{align}
The irreducible representations have the
following fusion property
\begin{align}
\cV^{U(1)_M}_k \otimes \cV^{U(1)_M}_{k'} = \cV^{U(1)_M}_{k+k'}, \ \ 
k,k' \in \Z_{2M} . 
\end{align}

\subsection{ $\frac{U(1)_M}{\Z_2}$-orbifold simple-current algebra}

The $\frac{U(1)_M}{\Z_2}$-orbifold simple-current algebra
$\cV^{\frac{U(1)_M}{\Z_2}}_1$ is generated by the spin-$M$ simple current
$\psi=\cos(\sqrt{2M}\phi)$.  Note that $\cV^{\frac{U(1)_M}{\Z_2}}_1$ is the
$\Z_2$ invariant part of $\cV^{U(1)_M}_1$, where $\Z_2$ acts as
\begin{align}
 \Z_2: \phi \to -\phi.
\end{align}
$\cV^{U(1)_M}_1$ contains $\cV^\text{Vir}_1$ generated by energy momentum
tensor $T \sim j^2$ which is $\Z_2$ invariant.  Thus
$\cV^{\frac{U(1)_M}{\Z_2}}_1$ also contains $\cV^\text{Vir}_1$, and $T$ acts
within $\cV^{\frac{U(1)_M}{\Z_2}}_1$.  $\cV^{U(1)_M}_1$ also contains
$\cV^{U(1)}_1$ which contains $\cV^\text{Vir}_{c=1,n^2},\ n \in \N$.  But the
states in $\cV^\text{Vir}_{c=1,n^2}$ transform as $|\psi\>\to (-)^n |\psi\>$
under the $\Z_2$.  So $\cV^{\frac{U(1)_M}{\Z_2}}_1$ only contains
$\cV^\text{Vir}_{c=1,n^2}$ for $n$ even:  
\begin{align}
 \cV^{U(1)}_1 = \bigoplus_{n\geq 0, n\ \text{even}} \cV^\text{Vir}_{c=1,n^2}
\end{align}
In particular, $j_4$ acts within
$\cV^{\frac{U(1)_M}{\Z_2}}_1$.

Now let us consider irreducible representations of the
$\frac{U(1)_M}{\Z_2}$-orbifold simple-current algebra.  We note that the $\Z_2$
action on the irreducible representations of the extended $U(1)_M$ current algebra
is given by
\begin{align}
\cV^{U(1)_M}_k \to \cV^{U(1)_M}_{-k} = \cV^{U(1)_M}_{2M-k} .
\end{align}
For $k= 1,\cdots, M-1$, the $\Z_2$ acts within  $\cV^{U(1)_M}_k\oplus
\cV^{U(1)_M}_{2M-k}$.  The $\Z_2$ even part of $\cV^{U(1)_M}_k\oplus
\cV^{U(1)_M}_{2M-k}$ forms an irreducible representation of
$\frac{U(1)_M}{\Z_2}$-orbifold simple-current algebra, denoted as
$\cV^{\frac{U(1)_M}{\Z_2}}_{\phi_k}$.  The corresponding primary field is
given by $\phi_{k}=\cos(k \phi/\sqrt{2M})$.  We know that the quantum dimension
of the representation  $\cV^{U(1)_M}_{\pm k}$is equal to 1. Thus the quantum
dimension for $\cV^{\frac{U(1)_M}{\Z_2}}_{\phi_k}$ is equal to 2.  The $\Z_2$
odd part of $\cV^{U(1)_M}_k\oplus \cV^{U(1)_M}_{2M-k}$ does not form an
irreducible representation of the $\frac{U(1)_M}{\Z_2}$ algebra.

The $\Z_2$ acts within $\cV^{U(1)_M}_1$. The $\Z_2$ even part of
$\cV^{U(1)_M}_1$ forms an irreducible representation of
$\frac{U(1)_M}{\Z_2}$ algebra, denoted as $\cV^{\frac{U(1)_M}{\Z_2}}_{1}$.
The corresponding primary field is the identity $1$.  The $\Z_2$ odd part of
$\cV^{U(1)_M}_1$ also forms an irreducible representation of the
$\frac{U(1)_M}{\Z_2}$ algebra, denoted as $\cV^{\frac{U(1)_M}{\Z_2}}_{j}$.
The corresponding primary field is the current operator $j$.  We also have two
new irreducible representations of the $\frac{U(1)_M}{\Z_2}$ algebra from the
twisted sector that twists the current $j$. The corresponding representations
are denoted as $\cV^{\frac{U(1)_M}{\Z_2}}_{\si^1}$ and,
$\cV^{\frac{U(1)_M}{\Z_2}}_{\tau^1}$.  The corresponding primary fields are
denoted as $\si^1$ and $\tau^1$.  We have the following fusion relations
for the irreducible representations
\begin{align}
 \cV^{\frac{U(1)_M}{\Z_2}}_{j}\otimes \cV^{\frac{U(1)_M}{\Z_2}}_{\si^1} &= \cV^{\frac{U(1)_M}{\Z_2}}_{\tau^1}, 
\nonumber\\
 \cV^{\frac{U(1)_M}{\Z_2}}_{j}\otimes \cV^{\frac{U(1)_M}{\Z_2}}_{\tau^1} &=
\cV^{\frac{U(1)_M}{\Z_2}}_{\si^1} .
\end{align}
Note that the irreducible representations correspond to the topological
excitations. The  fusion relations for the irreducible representations give
rise to the fusion relations of the  topological excitations.

Similarly, the $\Z_2$ acts within $\cV^{U(1)_M}_M$. The $\Z_2$ even part of
$\cV^{U(1)_M}_M$ forms an irreducible representation of $\frac{U(1)_M}{\Z_2}$
algebra, denoted as $\cV^{\frac{U(1)_M}{\Z_2}}_{\phi^1_M}$. The corresponding
primary field is $\phi^1_M=\cos(\sqrt{M/2}\phi)$.  The $\Z_2$ odd part of
$\cV^{U(1)_M}_M$ also forms an irreducible representation of
$\frac{U(1)_M}{\Z_2}$ algebra, denoted as $\cV^{\frac{U(1)_M}{\Z_2}}_{\phi^2_M}$.
The corresponding primary field is $\phi^2_M=\sin(\sqrt{M/2}\phi)$.  We also
have two new irreducible representations of the $\frac{U(1)_M}{\Z_2}$ algebra from
the twisted sector that twist $j$: $j\to -j$. The corresponding
representations are denoted as $\cV^{\frac{U(1)_M}{\Z_2}}_{\si^2}$ and,
$\cV^{\frac{U(1)_M}{\Z_2}}_{\tau^2}$. The corresponding primary fields are
denoted as $\si^2$ and $\tau^2$.  We have the following fusion relations
\begin{align}
 \cV^{\frac{U(1)_M}{\Z_2}}_{j}\otimes \cV^{\frac{U(1)_M}{\Z_2}}_{\si^2} &= \cV^{\frac{U(1)_M}{\Z_2}}_{\tau^2}, 
\nonumber\\
 \cV^{\frac{U(1)_M}{\Z_2}}_{j}\otimes \cV^{\frac{U(1)_M}{\Z_2}}_{\tau^2} &= \cV^{\frac{U(1)_M}{\Z_2}}_{\si^2}
.
\end{align}

\begin{table}
\caption{\label{U1/Z2rep} The irreducible  representations
$\cV^{\frac{U(1)_M}{\Z_2}}_{\al}$ of the $\frac{U(1)_M}{\Z_2}$-orbifold simple
current  algebra. The second column gives the conformal dimensions $h_\al$ of the
corresponding primary fields.  The third column are the quantum dimensions
$d_\al$ of the representations.  }
\begin{center}
\begin{tabular}{|ccc|c|}
\hline
label $\al$ & $h_\al$ & $d_\al$ &\\
\hline
 1& 0 & 1 &\\
$j$ & 1 & 1 & \\
$\phi_M^i$ & $M/4$ & 1 & $i=1,2$\\
$\sigma^i$ & 1/16 & $\sqrt M$ & $i=1,2$\\
$\tau^i$ & 9/16 & $\sqrt M$ & $i=1,2$\\
$\phi_k$ & $k^2/4M$ & $2$ & $k=1,\cdots,M-1$\\
\hline
\end{tabular}
\end{center}
\end{table}

The above are all the irreducible  representations of the 
$\frac{U(1)_M}{\Z_2}$-orbifold simple-current  algebra.  Table \ref{U1/Z2rep}
summarize the result.
\Ref{DVV8985} computed the full fusion rules of those irreducible
representations.  The fusion rules and conformal dimensions of the corresponding
primary fields are obtained by studying the modular transformation properties
of the characters.

It turns out that the $U_{1}(1)/\Z_2$ orbifold is the $U(1)_4$ Gaussian theory,
the $U_{2}(1)/\Z_2$ orbifold is two copies of the Ising CFT, and the $U_{3}(1)/\Z_2$
orbifold is the $\Z_4$ parafermion CFT of Zamolodchikov and
Fateev.\cite{ZF8515} 

\subsection{$U(1)_M$ topological orders}

The $U(1)_M$ simple-current algebra is generated by a single
simple-current operator $\psi=\ee^{\ii\sqrt{2M}\phi}$ (plus its hermitian
conjugate) with conformal dimension $h=M$.  The correlation function of $\psi$
gives rise to a bosonic Laughlin wavefunction
\begin{align}
 \prod_{i<j} (z_i-z_j)^{2M} = \< \psi(z_1) \psi(z_2) \psi(z_3)\cdots \> 
\end{align}
as discussed in Section \ref{corr}.
The simple-current primary fields
$\si_\al =\ee^{\ii \frac{\al }{\sqrt{2M}}\phi}$, $\al=1,\cdots,2M-1$,
produce wavefunctions that contain excitations with non-trivial topological type
\begin{align}
 \prod_i (\xi-z_i)^\al \prod_{i<j} (z_i-z_j)^{2M} = 
\< \si_\al(\xi) \psi(z_1) \psi(z_2)\cdots \>
\end{align}
We see a one-to-one correspondence between the simple-current primary fields
and the topological excitations.

The above picture is valid even when $M$ is half-integer. In this case the
correlation function of $\psi$ gives rise to a fermionic Laughlin wavefunction,
and the simple-current primary fields $\si_\al =\ee^{\ii \frac{\al
}{\sqrt{2M}}\phi}$, $\al=1,\cdots,2M-1$, give rise to the topological
excitations in the fermionic Laughlin state.

\subsection{$U(1)_M/\Z_2$-orbifold topological orders}

The $U(1)_M/\Z_2$-orbifold simple-current algebra is generated by a single
simple-current operator $\psi=\cos(\sqrt{2M}\phi)$ with conformal dimension
$h=M$. We note that $\psi^2\sim 1$ (\ie the OPE of two $\psi$'s produces the
identity operator $1$ as the leading term).

The correlation function of $\psi$'s 
\begin{align}
 \Psi(\{z_i\}) \propto \lim_{z_\infty\to \infty} \< \hat V(z_\infty) \prod  
\psi(z_i) \>
\end{align}
is single-valued (no branch cut) since the conformal dimension of $\psi$ is
integer and the OPE of $\psi$'s only produces operators with integer conformal
dimensions.  Also, since $\psi$ has an integer conformal dimension and is bosonic,
the correlation function $\Psi(\{z_i\})$ is a symmetric function, which gives
rise to a quantum Hall many-boson wavefunction $\Psi(\{z_i\}) \ee^{-\frac14
\sum |z_i|^2}$ with a bosonic topological order.  The edge excitations of such
a quantum Hall state are described by the $\frac{U(1)_M}{\Z_2}$-orbifold CFT,
the CFT that produces the bulk wave function, as calculated in
\Ref{WWH9476,W9927,LWW1024,Wtoprev}.

However, the above construction has a problem: the correlation of $\psi$'s (\ie
$\Psi(\{z_i\})$) has poles as $z_i\to z_j$.  But this is only a technical
problem that can be fixed.  We may put the wave function on a lattice or add
additional factors, such as $\prod |z_i-z_j|^{2M}$, to make the wave function
finite.  

We may also combine the $\frac{U(1)_M}{\Z_2}$-orbifold simple-current algebra with
a Gaussian model, as described in Section \ref{FQHVA}, to produce a many-body
wave function without poles.  We can choose the Gaussian model to have two
fields $\v \phi=(\phi^1,\phi^2)$ and choose $G^{\mu\nu}$ to be
\begin{align}
 G= \bpm 
2M & 1 \\
1 & 0 \\
\epm .
\end{align}
We choose the three simple currents as
\begin{align}
 c_1 = \psi \ee^{\ii \phi^1},\ \ \ 
 c_2 = \ee^{\ii \phi^2}, \ \ \ 
 c_3 = \ee^{\ii \phi^1} ,
\end{align}
which corresponds to choosing $k^I_\mu$ as
\begin{align}
k^I_\mu= \bpm 
1 & 0 \\
0 & 1 \\
1 & 0 \\
\epm_{I\mu} .
\end{align}
The $2M^\text{th}$ order zero in the correlation function of $\ee^{\ii \phi^1}$
cancels the $2M^\text{th}$ order pole in the correlation function of the $\psi(z_i)$. So
the correlation functions of $c_I$, $I=1,2,3$, are single-valued and finite,
which gives rise to a triple-layer bosonic wavefunction:
\begin{align}
& P(\{z_i,w_i,u_i\}) =
\\
&
\ \< c_1(z_1) c_1(z_2) \cdots c_2(w_1) c_2(w_2) \cdots c_3(u_1) c_2(u_2) \cdots \> .
\nonumber 
\end{align}

To understand the topological excitations in such a triple-layer state, we
note that $c_I$-primary fields have the form
\begin{align}
\eta_{\al,\v l^\al} 
= \si_\al \ee^{\ii \v l^\al\cdot \v\phi},\ \ \  \v\phi =(\phi^1,\phi^2),
\end{align}
where $\v l^\al$ satisfies \eqn{Klh}.
Since
$ \hsc{\al,\vec a}-   \hsc{\vec a} -\hsc{\al}$ are integers for all $\vec a$,
we find that the $\v l^\al$ satisfy
$ a^I k^I_\mu G^{\mu\nu} l^\al_\nu \in \Z$ 
or
\begin{align}
\bpm 
2M & 1 \\
1 & 0 \\
2M & 1 \\
\epm 
\bpm 
l^\al_1\\
\l^\al_2
\epm
=
\bpm 
0 \text{ mod } 1\\
0 \text{ mod } 1\\
0 \text{ mod } 1\\
\epm .
\end{align}
The above requires $\v l^\al$ to be integer vectors, and all the different $\v
l^\al$ are equivalent. So we can choose $\v l^\al=0$.  

We would like to remark that if we did not include the simple current
$c_3$ for the third layer, $\ee^{\ii \phi^1}$ would correspond to a non-trivial
primary field which would lead to extra topological types. With the  simple
current $c_3$, $\ee^{\ii \phi^1}$ will be a descendent field of the simple
current algebra, and will not correspond to a new type of topological
excitation.  We also like to remark that there is no particle number
conservation, for each layer or for all the layers.  If we did have  particle
number conservation for each layer, the constructed state may spontaneous break
such particle-number-conservation symmetry and contain gapless Goldstone modes.

We note that $G^{\mu\nu}$ has negative eigenvalues and the corresponding purely
chiral CFT is not unitary.  This can be fixed by treating the part of
$G^{\mu\nu}$ with negative eigenvalues as anti-holomorphic (\ie producing
correlations that depend on $z^*$).  We may also remove the poles using purely
chiral unitary CFT that describes the $E_8$ quantum Hall state, \ie using
eight scalar fields $\phi^i$, $i=1,\cdots,8$ and choosing
\begin{align} 
G={\footnotesize \begin{pmatrix}
2&1&0&0&0&0&0&0\\ 
1&2&1&0&0&0&0&0\\ 
0&1&2&1&0&0&0&0\\ 
0&0&1&2&1&0&0&0\\
0&0&0&1&2&1&0&1\\ 
0&0&0&0&1&2&1&0\\ 
0&0&0&0&0&1&2&0\\ 
0&0&0&0&1&0&0&2\\
\end{pmatrix} }  ,
\end{align}
to form
nine simple-current operators
\begin{align}
c_i =\ee^{\ii \phi^i}|_{i=1,\cdots,8},\ \ \ 
c_9 =  \psi \ee^{\ii \phi^1}.
\end{align}
This can remove the pole for the $M=1$ case.  To remove the pole for other
cases with larger $M$, we can add several copies of $E_8$ quantum Hall states.
The new simple-current algebra has the same topological excitations as the
$\frac{U(1)_M}{\Z_2}$-orbifold simple-current algebra, and has the same central
charge mod 8.  In this paper, we will use Gaussian theory with $G^{\mu\nu}$
that may have negative eigenvalues to remove the poles.  We can also choose the
Gaussian theory to be several copies of $E_8$ states to remove the poles.

We see that the topological excitations in our triple-layer bosonic wave
function are in one-to-one correspondence with the primary fields (or the
irreducible representations) of the $U(1)_M/\Z_2$-orbifold simple-current
algebra.  The chiral central charge of our triple-layer bosonic state is $c=1$ 
(1 from the  $U(1)_M/\Z_2$ simple-current and $1+(-1)$ from the Gaussian CFT).  
The $U(1)_M/\Z_2$ order is of type $N^B_c=(7+M)^B_1$.

\subsection{Reduction to smaller $N$}

We now establish that for $M$ odd, the topological order $U(1)_M/\Z_2$ can be reduced as
\begin{align}
& M=4p+3: 
\nonumber\\
 &\quad (N,c)=(7+M, 1) \rightarrow (N,c)=({7+M \over 2}, 2),
\nonumber \\
& M=4p+1: 
\nonumber \\
&\quad  (N,c)=(7+M,1) \rightarrow (N,c)=({7+M \over 2},0).
\label{eq:halfOrbi}
\end{align}
This reduction is similar in spirit to the reduction $(A_1,k) \rightarrow
(A_1,k)_{1 \over 2}$ which we discuss in section \ref{sec:su2red}. In the
Tables \ref{toplst5}-\ref{toplst7} we marked these reduced orders as $(U(1)_M/\Z_2)_{1 \over 2}$.

We first consider the case $M=3$. We already remarked that this order, with $N=10$ primaries, precisely agrees with the $\Z_4$ parafermions. The dictionary reads (see section \ref{sec:Zkpf} for notation)
\begin{align}
&& \phi_3^1 \rightarrow  \psi_1, \quad j \rightarrow  \psi_2, \quad \phi_3^2 \rightarrow  \psi_3,
\nonumber \\
&& \sigma^1 \rightarrow \Phi^1_1, \quad
\tau^1 \rightarrow \Phi^1_5, \quad
\sigma^2 \rightarrow \Phi^1_7, \quad
\tau^2 \rightarrow \Phi^1_3, 
\nonumber \\
&& \phi_1 \rightarrow \Phi^2_2, \quad
\phi_2 \rightarrow \Phi^2_4.
\end{align}
Following standard practice (see section \ref{sec:su2red}) we can now combine these fields with a single scalar field so as to produce the current 
algebra for $SU(2)_4$ at $c=2$. As explained in section \ref{sec:KMcurrents} this current algebra gives rise to $k+1=5$ primary sectors. For example, the sector with $s={1 \over 8}$, $d=\zeta_4^1$ comprises the fields
\begin{align}
(\sigma^1 \ee^{\ii {1 \over \sqrt{8}} \phi}, \ 
\tau^2 \ee^{\ii {3 \over \sqrt{8}} \phi}, \
\tau^1 \ee^{\ii {5 \over \sqrt{8}} \phi}, \ 
\sigma^2 \ee^{\ii {7 \over \sqrt{8}} \phi}).
\end{align}
We thus establish that the order $N^B_c=5^{B,a}_2$ in Table \ref{toplst5} is generated by the CFT $(U(1)_3/\Z_2)_\frac12$. 

This construction of the order $(U(1)_3/\Z_2)_\frac12$ is an example of a simple-current reduction of the product of two topological orders. The building blocks are the orders $(U(1)_3/\Z_2)$, with $N=10$, $c=1$, and $U(1)_4$, with $N'=8$, $c'=1$. In the product theory we can define the bosonic simple currents 
\begin{align}
1,\quad \phi_3^1 \ee^{\ii {1 \over \sqrt{2}} \phi} ,\quad j \ee^{\ii {2 \over \sqrt{2}} \phi},\quad \phi_3^2 \ee^{\ii {3 \over \sqrt{2}} \phi}\ .
 \end{align}
Of the $N \times N' = 80$ fields in the product theory, 20 are local with respect to all bosonic simple currents. These fields organize into 5 orbits and make up a reduced order of rank $N\times N'/16 =5$  and central charge $c+c'=2$. In formula
\begin{align}
(U(1)_3/\Z_2)_\frac12 = [ U(1)_3/\Z_2 \otimes U(1)_4 ]_{1 \over 16} .
\end{align}

Turning to $M=5$, we can follow a similar logic, but with an important twist: the scalar field now comes with metric $G=-1$, implying that it contributes $c=-1$ to the total central charge, and that a vertex operator $\ee^{\ii a \phi}$ has conformal dimension $s=-{a^2 \over 2}$. In section \ref{sec:su2red} we see similar minus signs in the construction of $(A_1,k)_{1 \over 2}$ for $k=4p+1$.
We can define a set of bosonic currents according to 
\begin{align}
1,\quad \phi_5^1 \ee^{\ii {1 \over \sqrt{2}} \phi} ,\quad j \ee^{\ii {2 \over \sqrt{2}} \phi},\quad \phi_5^2 \ee^{\ii {3 \over \sqrt{2}} \phi}\ .
 \end{align}
With respect to these currents, the following field combinations are primary and mutually inequivalent
\begin{align}
1, \quad j, \quad  \phi_2, \quad
\phi_1 \ee^{\ii {1 \over \sqrt{2}} \phi}, \quad
\sigma^1 \ee^{\ii {1 \over \sqrt{8}} \phi}, \quad \sigma^1 \ee^{\ii {3 \over \sqrt{8}} \phi},
\end{align}
with $s=0, 0, {1 \over 5}, -{1 \over 5},  0, {1 \over 2}$ and $d=1, 1, 2, 2,
\sqrt{5}, \sqrt{5}$. We thus recover the $6^{B,a}_0$ topological order in Table \ref{toplst6}.
The pattern for general odd $M$ is similar and leads to the result given in (\ref{eq:halfOrbi}).

\section{2+1D topological orders from Kac-Moody current algebra}
\label{sec:KMcurrents}

A rich class of simple-current algebras in CFT is provided by the affine
Kac-Moody algebras $X_l^{(1)}$ at positive integer level $k$. To each choice
$(X_l,k)$ corresponds a unitary CFT (the level-$k$ WZW model on the associated
group manifold) whose current algebra consists of currents $J^A(z)$, with
$A=1,2,\ldots, D$ an adjoint index of the Lie algebra $X_l$. The central charge
of this CFT is
\begin{equation}
c(X_l,k) = {k D \over k+g}
\end{equation}
with $D$ the dimension of $X_l$ and $g$ the dual Coxeter number. We provide some details in Appendix \ref{KM}, where we have also tabulated $(D,g)$ for the simple Lie algebras $X_l$. 

Starting from topological orders of Kac-Moody type, one may look for additional
bosonic simple currents and use these to extend the bosonic simple-current
algebra. In some special cases, the CFT $(X_l,k)$ contains Kac-Moody primaries
that are bosonic simple currents, and the extended current algebra leads to a
novel type of topological order with reduced rank $N$. These orders are closely
related to exceptional modular invariant partition functions (MIPF) based on
these same simple currents \cite{SY90}. Examples are the orders $(A_1,k)_{1
\over 4}$ for $k=4,8,\ldots$ and $(A_2,k)_{1 \over 9}$ for $k=3,6,\ldots$,
which we present below.

A more general, but often simpler, case involves the addition of one or several
scalar fields (or $U(1)$ factors) and the use of simple currents of the form
\begin{align}
c_{I,\v k^I} = \psi_I V_{\v k^I}
\end{align}
where the $\psi_I$ are simple currents in the $(X_l,k)$ CFT and the $V_{\v k^l}$ are scalar field vertex operators . Examples are reductions of type $(A_n,k)_{1 \over n+1}$ (see section \ref{sec:su2red}, \ref{sec:higherrank}) and the reductions
\begin{align}
& T_2: \
N^B_c \to [ N^B_c \otimes 4_0^{B,b}]_{1 \over 4} &
\label{eq:reductionstar}
\\ 
& T_8:  \ 
N^B_c \to [ N^B_c \otimes 4_1^B]_{1 \over 4} &
\label{eq:reductionstarstar}
\end{align}
discussed in sections \ref{sec:star}, \ref{sec:su2kstarstar} below.  Here
$4_0^{B,b}$ and $ 4_1^B$ are the bosonic topological orders in table
\ref{toplst}.  $ 4_0^{B,b}$ is the double-semion topological order and $ 4_1^B$
is $\nu=1/4$ bosonic Laughline state.

We note that both the operations $T_2$ and $T_8$ do not change the number $N$ of
topological types neither the quantum dimensions $d_i$.  The operation
$T_2$ also does not change the central charge $c$.  In
contrast, the operation $T_8$ changes the central charge by
$+1$.  Both the operations do change the spins $s_i$.  We also like to point
out that the operation $T_2$ is a $\Z_2$ operation, while the
operation $T_8$ is a $\Z_8$ operation.

Yet more general are cases where the additional bosonic simple currents contains factors in different non-Abelian orders. One example is the case
\begin{align}
[ (C_4,1) \times (A_1,1) \times (A_3,1) ]_{1 \over 4}
\end{align}
which turns out to be equivalent to a CFT coset construction and gives rise to the $c={4 \over 5}$ minimal model of the Virasoro algebra, of rank $N=10$ (see section \ref{sec:coset}).

In these constructions, it is sometimes convenient to first pass from the
$(X_l,k)$ CFT to the (generalised) parafermion CFT \cite{Gep87} obtained by
modding out $U(1)^l$, and then use the parafermions $\psi_\La$, which are simple
currents, as building blocks in the construction of an (extended) bosonic
simple-current algebra.

We remark that the simple-current reductions that we study here correspond to the condensation of bosonic topological excitations
\cite{BS08080627,K13078244,HW13084673,ERB1301,LWW1414}.

We have observed that we can construct all topological orders collected in
tables \ref{toplst}-\ref{toplst7} from orders based on Kac-Moody current
algebra $(X_l,k)$ and $U(1)$ factors if we use
\begin{itemize}
\item
conjugation by time reversal symmetry, sending 
$$
c \to -c, \quad d_i \to d_i, \quad s_i \to -s_i,
$$
\item
stacking of topological orders,
\item
simple-current reductions of (combinations of) topological orders.
\end{itemize}

The conformal blocks of the bosonic simple currents $c_I(z_i)$ will, in general
contain both zeros and poles in the differences $(z_i-z_j)$. To define a
many-body bosonic wave function, one needs to cancel the poles. This can be
done by including additional scalar fields, in such a way that essential
topological data (central charge and quantum dimensions and spins of all
excitations) are not affected. We make this step explicit in the examples of
$U(1)_M/\Z_2$ and $(A_1,k)$ in sections \ref{orb} and \ref{su2k}, and will
assume that a similar step is always possible in other cases. With that, we
arrive at bosonic many-body wave-functions for all cases listed in Tables
\ref{toplst}-\ref{toplst7}.

\subsection{$SU(2)_k$ current algebra}
\label{su2k}

The case $(A_1,k)$, commonly denoted as $SU(2)_k$, gives a CFT of central charge $c={3k \over k+2}$. 

The weight and root lattices (see appendix \ref{KM}) have the following structure. Writing the fundamental weight as 
$\Lambda_1={1 \over 2} {\bf e}_1$, the single positive root is $\alpha_1=2\Lambda_1$ and the Weyl group has two
elements: the identity and the reflection $w_1: \Lambda_1 \rightarrow -\Lambda_1$. 
A general (integral, dominant) weight is $\Lambda=l\Lambda_1, \ l\in \N$, so the irreducible representations are labeled by $l$. 

At level $k$ there are  $k+1$ irreducible representations (or primary fields)
$\Phi_l$,  $l=0,\cdots,k$, with conformal dimension (spin)
\begin{align}
 s_l =\frac{l^2+2l}{4(k+2)}.
\end{align}
The modular $S$-matrix is found to be 
\begin{equation}
S_{ll'} \propto \sin[ {\pi \over k+2}(l+1)(l'+1)]
\end{equation}
and the quantum dimensions are
\begin{equation}
d_l = {S_{0l} \over S_{00}} =  {  \sin[ { \pi \over k+2}(l+1)]  \over   \sin[{\pi \over k+2} ] } = \zeta^l_k \ .
\end{equation}

The $SU(2)_1$ Kac-Moody algebra is generated by three simple-current operators
$j^z,j^\pm$ with conformal dimension $h=1$.  In fact the $SU(2)_1$ algebra can be generated by a single simple current 
$j^+$ plus its hermitian conjugate.  
To obtain a many-body wave function without poles from the correlator of simple currents, we can combine the $SU(2)_1$ Kac-Moody 
algebra with a Gaussian model with two additional scalar fields $\v \phi=(\phi^1,\phi^2)$ with metric $G^{\mu\nu}$ given by
\begin{align}
 G= \bpm 
2 & 1 \\
1 & 0 \\
\epm ,
\end{align}
and choose the simple-current operators as
\begin{align}
 c_1 &= j^z \ee^{\ii \phi_1}, &
 c_2 &= j^+ \ee^{\ii \phi_1}, &
 c_3 &= j^- \ee^{\ii \phi_1}, 
\nonumber\\
 c_4 &= \ee^{\ii \phi_2}, &
 c_5 &= \ee^{\ii \phi_1}.  
\end{align}
We note that the 2nd order pole in the $j^z$-$j^z$ correlator is canceled by the
2nd order zero in the $\ee^{\ii \phi_1}$ correlator.  The finite correlators of the $c_I$ 
give rise to a (fractional) quantum Hall wavefunctions with 5 layers.  
We may also view the quantum Hall wavefunction  as a wave function in 3 layers, 
where the particles in the first layer carry spin-1.

For such choice of the Gaussian model, the  Gaussian model does not contribute
to chiral central charge, does not change the number of topological types, and
does not change the qauntum dimensions and spins of the topological excitations.  
The edge excitations of the constructed quantum Hall states are described by $SU(2)_k$
Kac-Moody algebra.

We see that the $N^B_c=2^B_1$ topological order in Table \ref{toplst} is described by 
$SU(2)$, $k=1$ Kac-Moody algebra, and we marked the entry as $(A_1,1)$. 
Similarly, we marked entries $(A_1,k)$ for orders $N^B_c$ given by $(k+1)^B_{3k /k+2}$, 
$k=2, \ldots 6$ in the corresponding tables. 

\subsection{Reductions of $SU(2)_k$ current algebra}
\label{sec:su2red}

A general affine Kac-Moody current algebra $X_l^{(1)}$ can be decomposed as a
product of (generalised) parafermions times a $U(1)^l$ scalar field factor
\cite{Gep87}.  For the case of $SU(2)_k $ this gives the familiar $\Z_k$
parafermions with central charge $c_k=2{k-1 \over k+2}$. The parafermions are
simple currents, but in general they are neither bosonic nor fermonic. 

In subsection \ref{sec:Zkpf} we briefly review $\Z_k$ parafermions and their
relation to $SU(2)_k$ current algebra.

Next we focus on orders $(A_1,k)_{1 \over 2}$ for $k$ odd, which contain half
the number of fields of $(A_1,k)$ and are realised at central charge
$c=c(A_1,k) \pm 1$. Our notation follows \Ref{RSW0777}.  We show how these
reduced orders arise through a simple-current reduction.

In subsection \ref{sec:su2kquarter} we present the orders $(A_1,k)_{1 \over 4}$
which employ a bosonic simple current that is part of the $SU(2)_k$ spectrum
for $k=4,8,\ldots$. A subtle point is the occurrence of `short orbits' of the
simple-current action, which lead to multiplicities in the modular invariants
\cite{SY90,BS08080627}.  The resolution of these multiplicities leads to novel modular
$S$-matrices, which are in general not captured by Kac-Moody current algebra
alone.  

\vskip 3mm

\subsubsection{$\Z_k$ parafermions and $(A_1,k)$ orders}
\label{sec:Zkpf}

The $\Z_k$ parafermion fields \cite{ZF8515} 
\begin{align}
 \psi_I, \ \ \ \ I=0,\cdots,k-1 ,
\end{align}
of conformal dimension $h_I={I(k-I) \over k}$, satisfy the operator algebra
\begin{align}
\psi_I(z) \psi_J(w) \sim (z-w)^{s_{IJ}} \psi_{I+J}
\end{align} 
with $s_{IJ}\equiv - {2 IJ \over k} \mod 1$. A general field in the parafermion theory is written as $\Phi^l_m$, 
$l=0,1,\ldots,k$ and $m \in \Z$, with conformal dimension
\begin {align}
s_{l,m} \equiv {l(l+2) \over 4(k+2)}-{m^2 \over 4k} \mod 1 .
\end{align}
The index $m$ is periodic with period $2k$ and $m \equiv l \mod 2$. In addition
we have the identification $\Phi^l_m = \Phi^{k-l}_{m+k}$. This leaves a total of ${k(k+1) \over 2}$ 
fields. All fields can be reached by acting with the parafermions
$\psi_i=\Phi^0_{2i}$ on the primaries $\sigma_i=\Phi^i_i$, $i=0,1,\ldots k$.
We also define $\epsilon_i = \Phi^{2i}_0$.

Using a single scalar field $\phi$ we can write the bosonic currents ($I=0,1,\ldots,k-1$, $j \in \Z$)
\begin{align}
 c_{I,j}=\psi_I \ee^{\ii k^{I,j} \phi} = \psi_I \ee^{\ii [{I \sqrt{2 \over k}} +  j \sqrt{2k}] \phi} ,
\end{align}
which have integer conformal dimension. The currents $c_{I=1,j=0}$ and $c_{I=k-1,j=-1}$ have conformal dimension 1. Together with $\ii \partial \phi$ they generate a level-$k$ affine Kac-Moody algebra $SU(2)_k$.
With respect to the bosonic chiral algebra $c_{I,j}$ the following fields represent admissible topological excitations
\begin{align}
\Phi^l_m \ee^{\ii [{m \sqrt{1 \over 2k}} + j \sqrt{2k}] \phi}
\end{align}
with $j \in \Z$. The excitations with $l=m=0,1,\ldots k$ and $j=0$ correspond to the highest weight states of the spin-$l$ representations of $SU(2)_k$. They constitute a set of $k+1$ inequivalent primaries of the bosonic current algebra. 

\subsubsection{The orders $(A_1,k)_{1 \over 2}$ with $k$ odd}
\label{sec:su2khalf}

In the $SU(2)_k$ theory, the field $\Phi_k$ is a simple current with fusion rules
\begin{align}
\Phi_{k} \Phi_l = \Phi_{k-l}, \quad l=0,1,\ldots k \ . 
\end{align}
This simple current can be used for a number of simple current reductions of the order $(A_1,k)$.

First assume that $k$ is odd and of the form $k=4p+3$. We can form a product with $U(1)_1 \sim SU(2)_1$, and 
consider the bosonic simple currents
\begin{align}
\Phi_0 \ee^{\ii {(2j) \over \sqrt{2}}\phi'}, \quad  \Phi_k \ee^{\ii {(2j+1) \over \sqrt{2}}\phi'}, \quad \quad j,j' \in \Z .
\end{align}
The primary sectors with respect to these currents are
\begin{align}
\Phi_l \ee^{\ii {(l+2j) \over \sqrt{2}}\phi'}, \quad l=0,1,\ldots {k-1 \over 2}, \quad j \in \Z.  
 \end{align}
They form the excitations of the reduced order $(A_1,k)_{1 \over 2}$ at $c=2{2k+1 \over k+2}$ and $N=(k+1)/2$. In formula we have
\begin{align}
(A_1,k)_{1 \over 2} = [ (A_1,k) \otimes U(1)_1 ]_{1 \over 4} , \quad k=3,7, \ldots
\end{align}
For $k$ of the form $k=4p+1$ one needs instead a factor $U(1)_1^* \sim SU(2)_1^*$ with $c=-1$ and non-trivial primary at $s=-{1 \over 4}$,
\begin{align}
(A_1,k)_{1 \over 2} = [ (A_1,k) \otimes U(1)_1^* ]_{1 \over 4} , \quad k=5,9, \ldots
\label{eq:su2khalfstar}
\end{align}

It is instructive to re-examine these same reductions starting from $\Z_k$ parafermions $\psi_I$, $I=0,\ldots k-1$, and the two scalar fields $\phi$, $\phi'$. For $k=3$ and with respect to the basis
\begin{align}
\phi_1 = \sqrt{2 \over 3} \phi, \quad \phi_2= \sqrt{1 \over 6} \phi + \sqrt{1 \over 2} \phi'
\end{align}
the metric becomes
\begin{align}
 G^{(3)}= \bpm 
{2 \over 3} & {1 \over 3} \\
{1 \over 3} & {2 \over 3}\\
\epm .
\end{align}
Writing $V_{\v k}$ for $\ee^{\ii \v k \cdot \v\phi}$, we can write bosonic currents
\begin{align}
c_{I,\v k^I} = \psi_I V_{\v k^I}
\end{align}
where 
$k_1^I,k_2^I$ are integers satisfying $2 k^I_1+k^I_2 \equiv 2I \mod 3$. The admissible topological excitations become
\begin{align}
\Phi^1_m  V_{\v k} \quad {\rm with} \quad  2k_1 + k_2 \equiv m \mod 3 \ .
\end{align}
Note that $\Phi^1_1=\sigma_1$, $\Phi^1_3=\epsilon_1$ and $\Phi^1_5=\sigma_2$. The fields $\Phi^1_mV_{\v k}$  form a single primary sector, with conformal dimension $s={2 \over 5}$ and quantum dimension $d=\zeta_3^1$, and we recover the order $(A_3,1)_{1 \over 2}=2^B_{14 \over 5}$.

For general $k=4p+3$, the 2-scalar metric becomes
\begin{align}
 G^{(k)}= \bpm 
{2 \over k} & {1\over k} \\
{1 \over k} & {1+k \over 2k} \\
\epm .
\label{eq:Gk}
\end{align}
The bosonic currents are
\begin{align}
c_{I,\v k^I} = \psi_I V_{\v k^I}
\end{align}
with $2k^I_1+ k^I_2 \equiv 2I \mod k$ and the primaries are 
\begin{align}
\Phi^l_m  V_{\v k} \quad {\rm with} \quad 2 k_1+ k_2 \equiv m \mod k ,
\end{align}
with $l=1,2,\ldots, {k-1 \over 2}$ and quantum dimensions $d=\zeta^l_k$. 

In this notation, the underlying $SU(2)_k \times SU(2)_1$ current algebra is formed by
\begin{align}
\psi_1 V_{(1\ 0)},\ \psi_{k-1} V_{(-1\ 0)}\ ; \quad V_{(-1 \ 2)},\ V_{(1 \ -2)};
\end{align}
together with the fields $\ii \partial \v \phi$. Odd-$l$ primaries under $SU(2)_k$ are doublets under the $SU(2)_1$, 
while even-$l$ primaries are singlets. 

For $k=3$ there is even more symmetry.  The following currents have conformal dimension equal to 1
\begin{align}
\psi_1V_{(1\ 0)},\  \psi_1V_{(-1\ 1)},\  \psi_1V_{(0\ -1)},\  
\nonumber\\
\psi_2V_{(-1\ 0)},\  \psi_2V_{(1\ -1)},\  \psi_2V_{(0\ 1)},\
\nonumber\\
V_{(2\ -1)},\ V_{(1\ 1)},\ V_{(-1\ 2)},\ 
\nonumber\\
V_{(1\ -2)},\ V_{(-1\ -1)}, V_{(-2\ 1)} . 
\end{align}
Together with $\ii \partial \v \phi$ these form the (14-dimensional) current algebra of $G_2^{(1)}$. 
The excitations, all of conformal dimension $s={2 \over 5}$,
\begin{align}
&& \epsilon_1,\ \sigma_1V_{(-1\ 0)},\  \sigma_1V_{(1\ -1)},\  \sigma_1V_{(0\ 1)},\  
\nonumber\\
& & \sigma_2V_{(1\ 0)},\  \sigma_2V_{(-1\ 1)},\  \sigma_2V_{(0\ -1)}\ 
\end{align} 
form the 7-dimensional representation of $G_2$.  Thus, the $(G_2,1)$ simple
current algebra can also produce the topological order $2^B_{14 \over 5}$.

For $k=4p+1$ the 2-scalar metric can be picked as 
\begin{align}
G^{\prime (k)} = \bpm 
{2 \over k} & {1 \over k} \\
{1 \over k} & {1-k \over 2k} \\
\epm .
\label{eq:Gkprime}
\end{align}
Note that the metric $G^{\prime (k)}$ has determinant $\det G^{\prime (k)} =-{1 \over k}$, whereas  $\det G^{(k)} ={1 \over k}$. This implies that for $k=4p+1$ the 2-scalar sector adds $1+(-1)=0$ to the total central charge, in agreement with eq. (\ref{eq:su2khalfstar}). The currents 
 \begin{align}
V_{(-1 \ 2)},\ V_{(1 \ -2)},
\end{align} 
have conformal dimension $-1$ and generate the algebra $SU(2)^*_1$. 

\subsubsection{The orders $(A_1,k)_{1 \over 4}$, $k=4,8,\ldots$}
\label{sec:su2kquarter}

For $k=4p$ the simple current $\Phi_k$ is bosonic and can be added to the currents of the $SU(2)_1$ Kac Moody algebra. 
In this situation, there exists a modular invariant partition function, labeled as ${\cal D}_{{k \over 2}+2}$, which only features the even-$l$ primaries
(see e.g.  \Ref{DiFMaSe97})
\begin{align}
{\cal D}_{{k \over 2}+2}: \quad Z_k = \sum_{l=0,2,\ldots}^{k-4 \over 2} |\chi_l+\chi_{k-l}|^2 + 2 |\chi_{{k \over 2}}|^2 \ .
\end{align}
Corresponding to this partition function is a bosonic topological order with $N={k \over 4}+2$, which we denote as $(A_1,k)_{1 \over 4}$.

The quantum dimensions of the fields $\Phi_l$, $l=0,2, \ldots {k-4 \over 2}$ are simply $\zeta_k^l$. The theory features two fields $\Phi_{k \over 2}^{(1)}$ and $\Phi_{k \over 2}^{(2)}$, which need to be `resolved' in the modular $S$-matrix \cite{SY90,BS08080627}. The result is that the two fields share the total quantum dimension  $\zeta_k^{{k \over 2}}$, leading to twice a value ${1 \over 2} \zeta_k^{{k \over 2}}$.

This construction for $k=4$ reproduces the abelian order at $N^B_c=3^B_2$, while for $k=8$ we reproduce the order at $N^B_c=4^B_{12/5}$ (we used $\zeta^2_8=(\zeta_3^1)^2$ and $\zeta_8^4=2\zeta_3^1$).

The case ${\cal D}_{8}$ at $k=12$ gives $5^B_{18/7}$ with
\begin{align}
& d=1,\ {1 \over 2} \zeta_{12}^6,\ {1 \over 2} \zeta_{12}^6,\ \zeta_{12}^2,\  \zeta_{12}^4  
\nonumber \\
& s=0,\ -{1 \over 7},\ -{1 \over 7},\ {1 \over 7},\ {3 \over 7} .
\end{align}
Using $\zeta_5^2 = {1 \over 2} \zeta_{12}^6$ we find a perfect match with the entry in the Table \ref{toplst5}.


Similarly, the entry at $6^B_{8/3}$ in Table \ref{toplst6} is found to agree
with the order $(A_1,16)_{1 \over 4}$. Note that $\zeta_7^3={1 \over 2}
\zeta_{16}^8 = \zeta_{16}^2$, revealing a triple degeneracy in the primary
sectors. This hints at an alternative interpretation, which we obtain in
section \ref{sec:A2oneninth}.

\subsubsection{The $\Z_8$ operation $T_8$ for $(A_1,k)$, $k=2,6,\ldots$}
\label{sec:su2kstarstar}

Inspecting the case $k=2,6,\ldots$, we find that the simple current $\Phi_k$
gives rise to yet another type of simple-current reduction. In this case, an
appropriate scalar field factor is $U(1)_2$, which is the order $4^B_1$.
Constructing the order
\begin{align}
[ (A_1,k) \otimes U(1)_2 ]_{1 \over 4} , \quad k=2,6, \ldots
\label{eq:su2halfstar} \end{align} we arrive at $N=k+1$, $c={3k \over k+2}+1$,
whereas the starting point $(A_1,k)$ corresponded to $N=k+1$, $c={3k \over
k+2}$. This reduction is thus an example of the operation $T_8$, which we defined in more 
general terms in the equation (\ref{eq:reductionstarstar}).

\subsection{Affine Kac-Moody algebras of higher rank} \label{sec:higherrank}

We can repeat the analysis for the $SU(2)$ case for the affine Kac-Moody
extension $X_l^{(1)}$ of all simple Lie algebras. As is well known, these have
been classified as four regular series $A_l$, $B_l$, $C_l$, and $D_l$,
$l=1,2,\ldots$ plus five exceptional algebras $E_6$, $E_7$ and $E_8$, $F_4$ and
$G_2$. This leads to many more examples of bosonic orders of low rank, which we
have marked in the tables. Note that $C_2\sim B_2$, $D_2\sim A_1\times A_1$,
$D_3\sim A_3$ and . In the tables we have displayed $c$ modulo 8 and conformal
dimensions $s_i$ modulo 1. 

A tentative list of simple-current primaries in the $X_l^{(1)}$ Kac-Moody
current algebras has been given in \cite{SY90}. As for the $SU(2)_k$ case,
these give rise to a variety of simple-current reductions of the order
$(X_l,k)$. 

For $(A_n,k)$ a reduction by a factor $\Z_{n+1}$ is possible if ${\rm
g.c.d.}(n+1,k)=1$ (see \Ref{RSW0777}), leading to orders $(A_n,k)_{1 \over
n+1}$. Below we discuss the cases with $n=2$ and the general case with level
$k=2$ and $n$ even. We remark that other reductions involving additional
abelian factors are possible, such as a reduction $(A_3,2)_{1 \over 2}$ which
leads to the order $5^{B,a}_{-2}$.

A second class are reductions based on bosonic simple-current primaries. Below
we present the case of $(A_2,k)_{1 \over 9}$.

\subsubsection{$(A_2,k)_{1 \over 3}$ for $k=2,4,5,7,\ldots$}

For $k=3p+2$ this reduction can concisely be written as 
\begin{align}
(A_2,k)_{1 \over 3} = [ (A_2,k) \otimes (A_2,1) ]_{1 \over 9} , \quad k=2,5, \ldots
\end{align}
For $k=2$ this reduces the order $(A_2,2)=6^B_{16/5}$ to $(A_2,2)_{1 \over 3}=2^B_{-14/5}$.  

One can re-examine this reduction in terms of the $SU(3)_2$ parafermions and four scalar fields.
For $k=3p+2$ the scalar field metric reads, in a convenient basis
\begin{align}
G^{(k)}= {1 \over k} \bpm 
2 & 1 & 2 & 2 \\
1 & 2 & 2 & 3 \\
2 & 2 & 4+2p & 4+p \\
2 & 3 & 4+p & 6+ 2p \\
\epm .
\label{eq:GmetricA2}
\end{align}
The currents 
\begin{align}
\Phi^{(0\ 0)}_{(\pm 2\ \mp 1)}V_{(\pm 1\ 0\ 0\ 0)},
\nonumber\\  
\Phi^{(0\ 0)}_{(\mp 1\ \pm 2)}V_{(0\ \mp 1\ 0\ 0)},\  
\nonumber\\
\Phi^{(0\ 0)}_{(\pm 1\ \pm 1)}V_{(\pm1\ \mp1\ 0\ 0)},
\end{align}
together with two scalars $\ii \partial \v \phi$, form an $SU(3)_k$ current algebra. In addition,
\begin{align}
V_{(\mp 1\ 0\ \pm 2\ \mp 1)},\  
V_{(0\ \mp 2\ \mp 1\ \pm 2)},\  
V_{(\mp 1\ \mp 2\ \pm 1\ \pm 1)}
\end{align}
together with the other two scalars form an $SU(3)_1$.

For $k=2$ the lattice defined by the matrix $G^{(2)}$ admits a total of  8 `short' integral vectors ${\v k}^S_i$, with ${\v k}^S_i \cdot {\v k}^S_i=1$, as well as 24 `long' integral vectors ${\v k}^L_i$, with ${\v k}^L_i \cdot {\v k}^L_i=2$. In fact, one recognizes in $G^{(2)}$ the metric of the $SO(9)$ weight lattice. Combining the integral vectors with a single Ising fermion (which is the parafermion for $SO(9)_1$), one can write a total of $24+8+4=36$ bosonic currents, which form the $SO(9)_1$ Kac-Moody current algebra. Combining these same vectors with the $SU(3)_2$ parafermions, which include three fields of conformal dimension $s={1 \over 2}$, leads to a total of $24+3\times 8+4=52$ bosonic currents, which form the Kac-Moody algebra for $F_4^{(1)}$ at level 1. Combining these same vectors with the $SU(3)_2$ parafermion spin fields, one can construct 26 fields of dimension $s={3 \over 5}$, which form an irreducible representation under $F_4$ and together constitute the single non-trivial primary sector of the topological order $N^B_c=2^B_{-14/5}$. 

For $k=3p+1$, the reduction becomes
\begin{align}
(A_2,k)_{1 \over 3} = [ (A_2,k) \otimes (A_2,1)^* ]_{1 \over 9} , \quad k=4,7, \ldots
\end{align}
We checked that for $k=4$ the quantum dimensions and spins of this reduced order match with the entry $N^B_c=5^B_{18/7}$ in Table \ref{toplst5}.

\subsubsection{$(A_n,2)_{1 \over n+1}$ for $n=2,4, \ldots$}

For $n=2,6,\ldots$, this reduction can be implemented as 
\begin{align}
(A_n,2)_{1 \over n+1} = [ (A_n,2) \otimes (\phi_1, \phi_2)]_{1 \over n+1} , \quad n=2,6, \ldots
\end{align}
with the scalar field metric given by (\ref{eq:Gk}) with $k=n+1$. The field content becomes
\begin{align}
\phi_{(l_1\ l_2\ \ldots \ l_n)} V_{\v k}
\end{align}
where the $l_j$ are the Dynkin labels of the $A_n$ representation carried by $\phi_{(l_1\ l_2\ \ldots \ l_n)}$ and 
\begin{align}
2k_1+ k_2 \equiv \sum_{j=1}^n j l_j \mod n+1 . 
\end{align}
This reduction adds $+2$ to the central charge. For  $n=4,8,\ldots$, one uses instead the metric (\ref{eq:Gkprime}) and the central charge remains unchanged.

We observe that there is a duality between the orders $(A_1,k)_{1 \over 2}$ and $(A_{k-1}, 2)_{1 \over k}$, in the sense that they form a pair $(N^B_c, N^B_{-c})$ with identical quantum dimensions $d_i$ and opposite spins $s_i$. This duality is a manifestation of the well-known level-rank duality between $SU(2)_k$ and $SU(k)_2$.

Other manifestations of level-rank duality are the pair $(A_1,4)$ and $(A_3,2)_{1 \over 2}$ and the pair $(A_2,4)_{1 \over 3}$ and $(A_3,3)_{1 \over 4}$, both with rank $N=5$.

\subsubsection{$(A_2,k)_{1 \over 9}$ for $k=3,6,\ldots $}
\label{sec:A2oneninth}

For $k=3p$, the $SU(3)_k$ primaries with weight $(k0)$ and $(0k)$ are
bosonic simple currents. They lead to an exceptional modular invariant, labeled ${\cal
D}_k$ in the classification of \Ref{Ga94}. These exceptional invariants only include fields with triality zero,
$l_1+2l_2 \equiv 0 \mod 3$. 

For $k=3$ the partition function is 
\begin{equation}
Z_3= |\chi_{(00)}+\chi_{(30)}+\chi_{(03)}|^2 +  3|\chi_{(11)}|^2.
\end{equation}
The corresponding order has 4 fields: the identity and 3 fields originating from $\phi_{(11)}$, with $d_i=1$, $s_i={1 \over 2}$. The value $d_i=1$ arises via equal distribution of the quantum dimension $d[\phi_{(11)}]=3$. We recognize the entry $N^B_c=4^B_4$.

For $k=6$, $c={16 \over 3}$, the modular invariant reads 
\begin{align}
Z= & |\chi_{(00)}+\chi_{(60)}+\chi_{(06)}|^2
\nonumber \\
& +  |\chi_{(11)}+\chi_{(41)}+\chi_{(14)}|^2
\nonumber \\
&  +  |\chi_{(33)}+\chi_{(30)}+\chi_{(03)}|^2 +  3|\chi_{(22)}|^2.
\end{align}
The weight $(2\, 2)$, with $s=-{1 \over 9}$, comes in with multiplicity 3 and quantum dimension $3 \zeta^3_{7}$ - after resolution into 3 primaries this leads to the values $d_i=\zeta^3_{7}$. The data for the other sectors are
\begin{align}
& (0\, 0),\ (6\, 0),\ (0\, 6): s=0, \ d=1
\nonumber \\
& (3\, 3),\ (3\, 0),\ (0\, 3): s= -{1 \over 3}, \ d= { \sin[{4 \pi \over 7}] \sin[{4 \pi \over 7}]  \over \sin[{\pi \over 7}] \sin[{2\pi \over 7}]} = \zeta^4_{16}
\nonumber \\
& (1\, 1),\ (4\, 1),\ (1\, 4): s={1 \over 3}, \ d= { \sin[{2 \pi \over 7}] \sin[{4 \pi \over 7}] \over \sin[{\pi \over 7}] \sin[{\pi \over 7}]} = \zeta^6_{16},
\end{align}
all in agreement with the data for the entry $6^B_{-8/3}$. 

For general $k=3p$, the rank of the order $(A_2,k)_{1 \over 9}$ is $N=(k^2+3k)/18+3$.

\subsubsection{The $\Z_2$ operation $T_2$}
\label{sec:star}

Inspecting the table \ref{toplst5} of rank-5 orders, we observe that $5^{B,a}_{\pm 2}$ derive directly from Kac-Moody current algebra, but $5^{B,b}_{\pm 2}$ do not. We remark that the orders $5^{B,b}_{\pm 2}$ arise through a simple-current reduction of the product of $5^{B,a}_{\pm 2}$ with $4_0^{B,b}$,
\begin{align}
5^{B,b}_{\pm 2} = [ 5^{B,a}_{\pm 2} \otimes 4_0^{B,b}]_{1 \over 4} \ .
\end{align}
This is a special case of the operation $T_2$ defined in \eqn{eq:reductionstar}. Similar doublets under the action of $T_2$ are 
$(6_0^{B,a},6_0^{B,b})$, $(6_4^{B,a},6_4^{B,b})$, and $(7_{\pm 2}^{B,b}, 7_{\pm 2}^{B,c})$.

\subsubsection{More general reductions}
\label{sec:coset}

We already mentioned that simple-current reductions of products of non-abelian
orders are possible. While these are not needed to reproduce the $N\leq 7$
orders that we list in this paper, they are needed to cover such cases as
minimal models of the Virasoro or ${\cal W}_n$ algebras, which are understood
via a coset construction \cite{BouSchou93,DiFMaSe97}. The idea is that a
coset $G/H$ is viewed as $G \times H^{-1}$ and that the corresponding order can
be obtained as a simple-current reduction of the product of orders $G$ and $H^*$.
As a concrete example, consider the coset
\begin{align}
{ SU(2)_3 \times SU(2)_1 \over SU(2)_4},
\end{align}
which describes the $c={4 \over 5}$ unitary minimal model of the Virasoro algebra, of rank $N=10$. Inspecting Table \ref{toplst5}, we see that the role of $(A_4,1)^*$ can be played by $(C_4,1)$. We therefore consider the product
\begin{align}
(C_4,1) \otimes (A_1,3) \otimes (A_1,1)
\end{align}
and pick as additional bosonic simple current the field
\begin{align}
\Phi_{(0001)} \times \Phi_3 \times  \ee^{\ii {\phi \over \sqrt{2}} } .
\end{align}
Of the $5\times 4\times 2 =40$ fields in the product theory, 20 are primary with respect to the extended simple-current algebra, and these organize into orbits of length 2. We thus recover the $N=10$ primary sectors of the minimal model.

\section{Summary}

In this paper, we use simple-current algebra to construct many-body wave
functions for 2+1D bosonic topological orders.  We found that simple-current
algebra can produce all the simple topological orders.  This supports the
conjecture that all the (non-)abelian statistics described by MTC can be
realized by bosonic systems. It also suggests that, in a certain sense,
simple-current algebra can be classified by MTC.

The simple-current reduction is an important tool in our constructions.  Such reductions 
correspond to the condensation of bosonic topological excitations.
\cite{BS08080627,K13078244,HW13084673,ERB1301,LWW1414} So the
simple-current reduction is also a tool to study the condensation of bosonic
topological excitations and the induced topological phase transition between
the original topological order and the reduced topological order.

\begin{acknowledgements}
KjS acknowledges hospitality at the Perimeter Institute, where part of this work was done.
Research at the Perimeter Institute is supported by the Government of Canada through Industry
Canada and by the Province of Ontario through the Ministry of Research.
The research of XGW is supported by NSF Grant No.~DMR-1005541 and NSFC 11274192, and
by the John Templeton Foundation No. 39901. The research of KjS is part of the Delta ITP consortium, 
a program of the Netherlands Organisation for Scientific Research (NWO) that is funded by the 
Dutch Ministry of Education, Culture and Science (OCW). 
\end{acknowledgements}

\appendix

\section{CFT of Kac-Moody current algebra}
\label{KM}

The starting point for the construction of a  CFT based on Kac-Moody current algebra is a simple Lie algebra 
$X_l$ plus a positive integer $k$ (which is called the level of the Kac-Moody current algebra).  
In this appendix we briefly review the connection between CFT and Kac-Moody current algebra and 
specify some of the data needed to identify key properties of the CFT.

\subsection{Root and weight lattices of finite dimensional Lie algebras}

In the structure theory of simple Lie algebras, it is common to choose a Cartan-Weyl basis $\{h^i, e_\alpha\}$, where the $h^i$, $i=1,\ldots,l$, form a basis of the Cartan sub-algebra $\cH$, and the $e_\alpha$ are ladder operators for the roots $\alpha=(\alpha^1, \ldots, \alpha^l)$,
\begin{equation}
 [h^i,e_\al]= \al^i e_\al .
\end{equation} 
The Killing form
\begin{equation}
K^{ab}=\Tr (\ad(J^a)\ad(J^b)),
\end{equation}
leads to an inner product in the root space $\cH^*$
\begin{equation}
 (\al,\bt)=\sum_{ij} K_{ij}\al^i\bt^j, \ \ \ \sum_j K_{ij}K^{jk}=\del_{ik}.
\end{equation} 

Integral linear combinations of the roots $\alpha$ form the so-called root-lattice associated 
with $X_l$. For a choice of simple roots $\alpha_i$, which form a basis of the root lattice, one defines 
the Cartan matrix is
\begin{equation}
A_{ij}=2\frac{(\al_i,\al_j)}{(\al_j,\al_j)} .
\end{equation}

Dual to the root lattice is the weight lattice, which plays a crucial role in a systematic description of the irreducible 
representations of $X_l$. Its elements $\La$ can be characterised by the Dynkin labels
\begin{equation}
l_i= 2 \frac{( \La,\al_i)}{(\al_i,\al_i)} .
\end{equation}
The weight is then written as a linear combination of fundamental weights $\La=\sum_i l_i \La_i$, 
where the fundamental weights have inner product
\begin{equation}
(\La_i, \La_j) = G_{ij}
\end{equation}
with 
\begin{equation}
G_{ij} = (A^{-1})_{ij} \frac{(\al_j,\al_j)}{2} .
\end{equation}

\subsection{Primaries of Kac-Moody current algebra}

The CFT associated with Lie algebra $X_l$ and level $k$ is characterised by a larger symmetry algebra, 
which is the so-called affine Kac-Moody extension or current algebra $X_l^{(1)}$ of $X_l$ at level $k$. 
The central charge of this CFT can be expressed as
\begin{align}
c = \frac{k D}{k+g}
\end{align}
where $D$ is the dimension of $X_l$ and $g$ is the dual Coxeter number. In table \ref{LieAlg} we list these data for 
the simple Lie algebras $X_l$.

The primary sectors of the current algebra CFT are labeled by particular weights $\La$ - the so-called dominant integral weights.
Their Dynkin labels satisfy $l_j\geq 0$ and 
\begin{align}
\sum_{j=1}^l l_j \, a_j^\vee \leq k,
\end{align}
where $a_j^\vee$ is the comark (or dual Kac label) to the root $\alpha_j$ \cite{GepWit86}.

The conformal dimension (spin) of the primary sector labeled by $\La$ is given by
\begin{equation}
 s_\La = \frac{(\La,\La+2\rho)}{2(k+g)}
\end{equation}
where $\rho=\sum_i \Lambda_i$ is the sum of the fundamental weights.

The $S$-matrix is given by
\begin{equation}
 S_{\La\La'} \propto \sum_{w\in W} \text{sign}(w) \ee^{-\frac{2\pi \ii}{k+g}
(w(\La+\rho),\La'+\rho)}
\end{equation}
where the summation is over the Weyl group of $X_l$. Via the relation 
\begin{equation}
d_i=\frac{S_{0i}}{S_{00}}
\end{equation}
this $S$-matrix fixes the quantum dimensions $d_i$.

\begin{table}
\caption{\label{LieAlg} Dimension $D$ and dual Coxeter number $g$ of the simple Lie algebras $X_l$.}
\begin{center}
\begin{tabular}{|cccc|}
\hline
algebra & & $D$ & $g$ \\
\hline
$A_l$ & $l\geq1$ & $l(l+2)$ & $l+1$ \\
$B_l$ & $l\geq2$ & $l(2l+1)$ & $2l-1$ \\
$C_l$ & $l\geq3$ & $l(2l+1)$ & $l+1$ \\
$D_l$ & $l\geq4$ & $l(2l-1)$ & $2l-2$ \\
$E_6$ & & 78 & 12 \\
$E_7$ & & 133 & 18 \\
$E_8$ & & 248 & 30 \\
$F_4$ & & 52 & 9 \\
$G_2$ & & 14 & 4 \\
\hline
\end{tabular}
\end{center}
\end{table}

We refer to \Ref{DiFMaSe97} for further details. Here, for the sake of illustration, we present such details for the rank-2 algebras $A_2$ (or $su(3)$), $B_2$ (or $so(5)$) and $G_2$.

\subsection{The rank 2 simple Lie algebras}

\subsubsection{The algebra $A_2$}

For this Lie algebra the weight-lattice metric $G_{ij}$ is given by
\begin{align}
 G= \bpm 
{2 \over 3} & {1 \over 3} \\
{1 \over 3} & {2 \over 3}\\
\epm .
\end{align}
With respect to an othonormal basis ${\bf e}_i$, the fundamental weights can be written as 
\begin{equation}
\Lambda_1= \sqrt{2 \over 3}  {\bf e}_2, \quad  \Lambda_2={1 \over \sqrt{2}} {\bf e}_1 +{1 \over \sqrt{6}} {\bf e}_2 ,  
\end{equation}
and the positive roots are 
\begin{align}
& \alpha_1 = -\Lambda_1 + 2 \Lambda_2, \quad \alpha_2 = 2\Lambda_1 - \Lambda_2,
\nonumber \\
& \alpha_{12} = \Lambda_1 + \Lambda_2,
\end{align}
which we write as $(-1\,2)$, $(2\,-1)$ and $(1\,1)$, respectively. The Weyl group has 6 elements, the orbit of $\rho=(1\ 1)$ is
\begin{eqnarray}
\text{sign}(w)=+1: & (1\ 1),\ (1\ -2),\ (-2\ 1)
\nonumber \\
\text{sign}(w)=-1: & (2\ -1),\ (-1\ 2),\ (-1\ -1) .
\end{eqnarray}
Dominant integral weights at level $k$ satisfy $l_1+l_2 \leq k$, their number is $N_k=(k+1)(k+2)/2$. The conformal and quantum dimensions for the primary $(l_1\ l_2)$ are given by 
\begin{align}
& s_{(l_1\ l_2)} = \frac{l_1^2 + l_2^2 + l_1 l_2 + 3l_1+3l_2}{3(k+3)},
\nonumber \\
& d_{(l_1\ l_2)} = \frac{ \sin[{\pi(l_1+1) \over k+3}] \sin[{\pi(l_2+1) \over k+3}] \sin[{\pi(l_1+l_2+2) \over k+3}] }{ \sin[{\pi \over k+3}] \sin[{\pi \over k+3}] \sin[{2\pi \over k+3}]} \ .
\end{align}
The central charges are $c={8k \over k+3}$ for the $SU(3)_k$ CFT and $c_k=6{k-1 \over k+3}$ for the corresponding parafermions.

\subsubsection{The algebra $B_2$}

For this Lie algebra the weight-lattice metric $B_{ij}$ is given by
\begin{align}
 G= \bpm 
1 & {1 \over 2} \\
{1 \over 2} & {1 \over 2}\\
\epm .
\end{align}
With respect to an othonormal basis ${\bf e}_i$, the fundamental weights can be written as 
\begin{equation}
\Lambda_1= {1 \over \sqrt{2}} {\bf e}_1 +  {1 \over \sqrt{2}} {\bf e}_2, \quad  \Lambda_2={\bf e}_1.
\end{equation}
The simple roots are 
\begin{align}
\alpha_1 = -\Lambda_1 + 2 \Lambda_2, \quad \alpha_2 = -2 \Lambda_1 + 2 \Lambda_2.
\end{align}
The four positive roots are
\begin{align}
-\Lambda_1 + 2 \Lambda_2, \ -2 \Lambda_1 + 2 \Lambda_2, \La_1,\  2 \La_2 \ .
\end{align}
The Weyl group has 8 elements, the orbit of $\rho=(1\ 1)$ is
\begin{eqnarray}
\text{sign}(w)=+1: & (1\ 1),\ (2\ -3),\ (-1\ -1), (-2\ 3) 
\nonumber \\
\text{sign}(w)=-1: & (-1\ 3),\ (-2\ 3),\ (1\ -3), (2\ -3) \ . 
\nonumber \\
\end{eqnarray}
Dominant integral weights at level $k$ satisfy $l_1+l_2 \leq k$. Their number is $N_k=(k+1)(k+2)/2$ and the conformal dimensions are given by 
\begin{align}
  s_{(l_1\ l_2)} = \frac{2 l_1^2 + l_2^2 + 2 l_1 l_2 + 6l_1+4l_2}{4(k+3)} \ .
\end{align}
The central charges are $c={10 k \over k+3}$ for the $SO(5)_k$ CFT and $c_k=2{k-3 \over k+3}$ for the corresponding parafermions.

\subsubsection{The algebra $G_2$}

The weight-lattice metric $G_{ij}$ is given by
\begin{align}
 G= \bpm 
2 & 1 \\
1 & {2 \over 3}\\
\epm .
\end{align}
With respect to an othonormal basis ${\bf e}_i$, the fundamental weights can be written as 
\begin{equation}
\Lambda_1= \sqrt{2} {\bf e}_2, \quad  \Lambda_2={1 \over \sqrt{6}} {\bf e}_1 +{1 \over \sqrt{2}} {\bf e}_2 ,  
\end{equation}
and the simple roots are 
\begin{align}
& \alpha_1 = -\Lambda_1 + 2 \Lambda_2, \quad \alpha_2 = 2 \Lambda_1 -3 \Lambda_2 .
\end{align}
The six positive roots are
\begin{align}
&  -\Lambda_1 + 2 \Lambda_2, \ 2 \Lambda_1 - 3 \Lambda_2,
\nonumber \\
& \La_1,\ \La_2,\ \La_1 - \La_2,\ -\La_1+ 3 \La_2 \ .
\end{align}
The Weyl group has 12 elements. The orbit of $\rho=(1\ 1)$ is
\begin{eqnarray}
w=+1: && (1\ 1),\ (-2\ 5),\ (-3\ 4), 
\nonumber \\ 
&& (-1\ -1),\ (2\ -5),\ (3\ -4)
\nonumber \\
w=-1: && (2\ -1),\ (-1\ 4),\ (-3\ 5), 
\nonumber \\ 
&& (-2\ 1),\ (1\ -4),\ (3\ -5).
\end{eqnarray}
Dominant integral weights at level $k$ satisfy $2l_1+l_2 \leq k$. Their conformal dimensions are given by
\begin{align}
  s_{(l_1\ l_2)} = \frac{3 l_1^2 + l_2^2 + 3 l_1 l_2 + 9 l_1+5 l_2}{3(k+4)} .
\end{align}
The central charges are $c={14k \over k+4}$ for the $(G_2)_k$ WZW model and $c_k=4{3k-2 \over k+4}$ for the corresponding parafermions.

%
%
%
%
%
%
%
%
%
%
%



\end{document}